\documentclass[floatfix,amssymb,preprint,aps,prl, superscriptaddress,showpacs]{revtex4}

\usepackage{amsmath}
\usepackage{amsfonts}
\usepackage{amssymb}
\usepackage{graphicx}
\usepackage{xcolor}
\usepackage{xspace}
\usepackage[normalem]{ulem}
\usepackage{textcomp}
\usepackage{newunicodechar}
\newunicodechar{ }{~}

\DeclareUnicodeCharacter{2009}{\,}
\DeclareUnicodeCharacter{2212}{-}

\begin{document}


\title{Percolative Instabilities and Sparse-Limit Fractality in 1T-TaS$_2$}

\author{Poulomi Maji}
\affiliation{School of Physical Sciences, Indian Association for the Cultivation of Science, 2A $\&$ B
Raja S. C. Mullick Road, Jadavpur, Kolkata - 700032, India
}

\author{Md Aquib Molla}
\affiliation{Vidyasagar College, Department of Physics, 39 Sankar Ghosh Lane, Kolkata 700006,India
}

\author{Koushik Dey}
\affiliation{School of Physical Sciences, Indian Association for the Cultivation of Science, 2A $\&$ B
Raja S. C. Mullick Road, Jadavpur, Kolkata - 700032, India
}

\author{Bikash Das}
\affiliation{School of Physical Sciences, Indian Association for the Cultivation of Science, 2A $\&$ B
Raja S. C. Mullick Road, Jadavpur, Kolkata - 700032, India
}
\author{Sambit Choudhury}
\affiliation{UGC-DAE Consortium for Scientific Research, Khandwa Road, Indore 452001, Madhya Pradesh, India}

\author{Tanima Kundu}
\affiliation{School of Physical Sciences, Indian Association for the Cultivation of Science, 2A $\&$ B
Raja S. C. Mullick Road, Jadavpur, Kolkata - 700032, India
}

\author{Pabitra Kumar Hazra}
\affiliation{School of Physical Sciences, Indian Association for the Cultivation of Science, 2A $\&$ B
Raja S. C. Mullick Road, Jadavpur, Kolkata - 700032, India
}

\author{Mainak Palit}
\affiliation{School of Physical Sciences, Indian Association for the Cultivation of Science, 2A $\&$ B
Raja S. C. Mullick Road, Jadavpur, Kolkata - 700032, India
}

\author{Sujan Maity}
\affiliation{School of Physical Sciences, Indian Association for the Cultivation of Science, 2A $\&$ B
Raja S. C. Mullick Road, Jadavpur, Kolkata - 700032, India
}
\author{Bipul Karmakar}
\affiliation{School of Physical Sciences, Indian Association for the Cultivation of Science, 2A $\&$ B
Raja S. C. Mullick Road, Jadavpur, Kolkata - 700032, India
}
\author{Kai Rossnagel}
\affiliation{Ruprecht Haensel Laboratory, Deutsches Elektronen-Synchrotron DESY, Notkestr. 85, 22607 Hamburg, Germany}
\affiliation{Institut für Experimentelle und Angewandte Physik, Christian-Albrechts-Universität zu Kiel, Olshausenstr. 40, 24098 Kiel, Germany}

\author{Sanjoy Kr Mahatha}
\affiliation{UGC-DAE Consortium for Scientific Research, Khandwa Road, Indore 452001, Madhya Pradesh, India}
\author{Bhaskaran Muralidharan}
\affiliation{Indian Institute of Technology Bombay, Mumbai - 400076, India}

\author{Shamashis Sengupta}
\affiliation{Université Paris-Saclay, CNRS/IN2P3, IJCLab, 91405 Orsay, France}

\author{Sanchari Goswami}
\affiliation{Vidyasagar College, Department of Physics, 39 Sankar Ghosh Lane, Kolkata 700006,India
}

\author{Subhadeep Datta}
\email{sspsdd@iacs.res.in}
\affiliation{School of Physical Sciences, Indian Association for the Cultivation of Science, 2A $\&$ B
Raja S. C. Mullick Road, Jadavpur, Kolkata - 700032, India
}

\begin{abstract}
\bfseries{ 
The low-temperature metallic phase of 1T-TaS$_2$ may originate from current- and voltage-driven destabilization of the commensurate charge density wave (CDW) in a strongly correlated Mott-insulator, alongside the robust yet rarely realized influence of intrinsic electronic distortions. Electrical pulse-driven transport, combined with second harmonic response, reveals abrupt switching, negative differential resistance (NDR), and multiscale domain-wall reorganization. The free energy analysis identifies a critical order parameter threshold for the Mott–metal transition, with scaling exponents ($\beta \!\approx\! 1.3$) consistent with 2D percolation. The sparse limit fractal dimension $D_f\!\approx\!0.3$ at 10~K, rising to $\approx\!0.9$ at 300~K, reflects the hierarchical evolution of the conductive pathways throughout the temperature. These findings establish a direct connection between fractal percolation, pulse-induced instabilities, and correlated electron transport, offering a framework for controlled access to non-equilibrium phase transitions in low-dimensional quantum materials.
}
\end{abstract}

\maketitle
\section{Introduction}

Electron-phonon interactions in transition metal dichalcogenides (TMDCs) are fundamental drivers of phase complexity, particularly through charge density wave (CDW) formation \cite{PhysRevLett.125.097002, zhu2015classification}. In metallic TMDCs, CDWs modulate the electron density ($\rho(r)$) periodically, coupling directly to lattice distortions \cite{PhysRevB.107.024101}. This intricate interplay produces a rich phase landscape with transitions across various commensurate and incommensurate phases, each stabilized by electron-phonon coupling ($g$). A comprehensive theoretical model for CDW behavior in TMDCs, however, remains incomplete. CDW formation is often described by Peierls instability, where lattice distortions (\( u(r) \)) lower electronic energy when \( 2k_F = q \), with the electron-phonon coupling represented by the Hamiltonian term \( H_{\text{CDW}} = g \sum_r (\rho(r) - \langle \rho(r) \rangle) u(r) \) \cite{PhysRevB.3.3173, PhysRev.167.691, friend1979periodic, rice1973theory}. Additional mechanisms like the Kohn anomaly and Fermi surface nesting have further refined this picture, but they do not yet encapsulate the full phenomenology observed in TMDCs \cite{PhysRev.131.1995, hoesch2009giant, PhysRevB.77.165135, rossnagel2011origin, aebi2001search, shen2008primary}.
On the transport side, Bardeen's quantum tunneling model suggests that CDW conduction can proceed through tunneling of the condensate across impurity-induced gaps \cite{bardeen1980tunneling}, while classical approaches treat depinning using a sinusoidal pinning potential and a single CDW phase degree of freedom. These models predict distinct nonlinear responses in voltage- versus current-driven regimes: voltage bias leads to a discontinuous rise in differential resistance beyond threshold, whereas current bias results in negative differential resistance (NDR) \cite{hall1984negative, van2001negative}. Such NDR, prominent in NbSe$_3$ and TaS$_3$, marks the threshold for abrupt collective CDW motion. However, these models lack an integrated understanding of criticality, fractal conductivity, and percolative instabilities intrinsic to domain evolution under pulses.

1T-TaS$_2$, a layered TMDC, hosts intertwined CDW and Mott insulating (MI) phases \cite{tsen2015structure, dardel1992spectroscopic}. Below 180~K, it stabilizes a commensurate CDW (C-CDW) with a $\sqrt{13} \times \sqrt{13}$ “Star of David” (SOD) superlattice \cite{mohammadzadeh2021room, yoshida2014controlling, wang2019lattice, stahl2020collapse,PhysRevB.100.155407,ritschel2015orbital}, accompanied by a MI state associated with “AL” stacking \cite{wu2022effect, wang2024dualistic, petkov2022atomic}. The nearly degenerate stacking of “L” promotes a metallic phase (ML), allowing co-existence and phase percolation \cite{lee2019origin, wu2022effect, ritschel2018stacking}. These interlayer arrangements modulate electronic bandwidth $W$ and its competition with Coulomb repulsion $U$, determining the emergent transport properties.

External stimuli -- optical pulses, gating, or dimensional confinement -- can tune interlayer hopping $t_\perp$ to overcome $U$, triggering the MI–ML transformation, which has been extensively investigated via local transport techniques such as scanning tunneling spectroscopy (STS) \cite{ma2016metallic, stahl2020collapse, sun2018hidden, vaskivskyi2015controlling, ishizaka2011femtosecond, wen2019photoinduced, han2015exploration, yu2015gate, yoshida2015memristive,vaskivskyi2015controlling}. Unlike smooth evolution in disordered films \cite{PhysRevB.35.197}, the transition in 1T-TaS$_2$ involves current-driven reorganization of the CDW domain, leading to abrupt, pulse-induced switching and NDR \cite{vaskivskyi2016fast, yoshida2015memristive,hollander2015electrically}. The emergence of sparse, metallic filaments follows a percolation-like dynamics governed by fractal geometry, local Joule heating, and electric-field focusing. The resistance scaling behavior is consistent with the modified Archie's law and the fractal connectivity models \cite{stauffer, PhysRevB.55.8038, Rabi}, revealing a new class of metastable, spatially inhomogeneous transport states. Despite the rich experimental observations, a quantitative framework describing the nonthermal fractal percolation kinetics and threshold-driven conductivity evolution in 1T-TaS$_2$ remains underexplored. This motivates our pulse-based approach, where we track transport transformations via $R$–$T$, $I$–$V$, and second harmonic response under varying current bias, providing insights into bistability, scaling behavior, and evolution of the order parameter throughout the MI–ML transition.

We report on transport in exfoliated and bulk 1T-TaS$_2$ flakes, showing an anomalous resistance drop near 100~K, deviating from conventional Mott insulating behavior. This transition, robust across thicknesses, suggests a stacking-induced instability that drives the system toward a metal-like (ML) state. Voltage-current (\textit{V-I}) measurements reveal negative differential resistance (NDR) from 10–300~K, with threshold current $I_{\mathrm{th}}(T)$ increasing up to 150~K, then decreasing, indicating current-driven percolative domain evolution. Short voltage or current pulses induce irreversible switching from the insulating towards ML state, highlighting the role of local Joule heating and field concentration near weak links. To describe the ML state's growth, we construct a phenomenological free energy $F[\phi]$ using a metallic order parameter $\phi$, representing volume fraction of conducting channels, with $0 \leq \phi \leq 1$, whose spatial profile evolves under both diffusive and gradient-driven terms. The interface width between domains depends on the ratio of diffusive strength to free energy curvature, allowing smooth transition to a percolative, conductive state as current increases. As $\phi \rightarrow 1$, sparse conductive backbones grow and merge, forming fractal pathways that abruptly reduce resistance beyond $I_{\mathrm{th}}$. The system’s bistability and nonlinearity emerge from this nonequilibrium, current-driven reorganization of electronic textures. These findings offer a unified framework for understanding current-induced insulator-to-metal transitions in correlated layered materials through spatially evolved percolation.

\section{Result and Discussion} 
1T-TaS$_2$ undergoes a series of charge-density-wave (CDW) transitions upon cooling, evolving from a metallic phase into an incommensurate CDW below $\sim550$~K and a nearly commensurate CDW (NC-CDW) below $\sim350$~K.
In the NC-CDW state, the lattice distortion condenses into hexagonal domains of Star-of-David (SoD) clusters (see Fig.~S1(b)) ~\cite{wang2019lattice, singh2022lattice, tsen2015structure}. Each SoD consists of 13 Ta atoms, where 12 peripheral atoms contract toward a central Ta, producing three inequivalent Ta sites. Upon further cooling below $\sim220$~K, these domains order into a long-range $\sqrt{13}\times\sqrt{13}$ superstructure, characteristic of the commensurate CDW (C-CDW) phase, in which only the central Ta atom hosts an unpaired $5d^1$ electron (see Fig.~S2(d)) \cite{mohammadzadeh2021room, Supple, wang2019lattice, stahl2020collapse,PhysRevB.100.155407,ritschel2015orbital}. The NC-CDW to C-CDW transition is clearly observed in Fig.~1(a) for both devices near $\sim220$~K for the heating cycle, and is further corroborated in Fig.~S3(b). The pronounced thermal hysteresis between the heating and cooling cycles in Fig.~S3(b) reflects the first-order nature of this transition \cite{yoshida2014controlling, tsen2015structure}. Although XPS does not resolve a clear distinction between the inequivalent B and C-type Ta atoms, the charge disproportionation between the central (A-type) and peripheral sites is manifested by a splitting of the Ta $4f_{7/2}$ and $4f_{5/2}$ core levels below 350~K (see Fig.~S1) \cite{salzmann2023observation, Supple, wang2024dualistic, kuhn2019directional}. The substantial electronic overlap among the peripheral Ta atoms is further supported by the calculated electron localization function of the supercell (Fig.~S2).
Below the C-CDW transition, the electronic ground state is governed by the out-of-plane stacking of the SoD clusters \cite{ritschel2018stacking, wu2022effect, petkov2022atomic}. In the AL configuration, vertical alignment of the central Ta atoms enhances the interlayer Coulomb repulsion $U$ and stabilizes a correlated Mott insulating state (Fig.~S1(b)) \cite{wu2022effect, wang2024dualistic, petkov2022atomic}. Deviations from this stacking destabilize the Mott phase and introduce metastability \cite{vaskivskyi2016fast, yoshida2015memristive,hollander2015electrically}. Moreover, scanning tunneling microscopy (STM) studies further show that voltage pulses can locally convert these insulating domains into more conducting regions by modifying the stacking order \cite{ma2016metallic, svetin2017three, stahl2020collapse}. First-principles calculations indicate that AL and L stackings are nearly degenerate, with the L stacking higher in energy by only $\sim 1.1$~meV per SoD \cite{lee2019origin, wu2022effect, ritschel2018stacking}. In the L configuration, lateral displacement of the clusters enhances both in-plane and interlayer hopping, yielding a metallic state\cite{lee2019origin,Supple,park2023stacking,wu2022effect}. This near-degeneracy provides a natural route to the Mott--metal transition observed under optical and ionic-liquid gating \cite{stahl2020collapse, sun2018hidden, vaskivskyi2015controlling, ishizaka2011femtosecond, wen2019photoinduced, han2015exploration, yu2015gate, yoshida2015memristive,vaskivskyi2015controlling}.
In our measurements (Fig.~1), signatures of Mott--metal coexistence emerge below $\sim 100$~K. While one device exhibits a sharp resistance drop consistent with an L-stacking-dominated metallic phase (indicated in black), another (indicated in red) shows a resistivity dip between 50--100~K accompanied by a broad, Gaussian-like second-harmonic resistance peak. In contrast to the sharp peak associated with the NC--CDW to C--CDW transition, this broadened second-harmonic response signals the low-temperature coexistence of competing Mott insulating and metallic phases, reflecting a continuous crossover from a mixed AL--L regime toward the AL-stacked Mott insulator, as shown in Fig.~1(a) inset and Fig.~S4(a) \cite{Supple}. Thus, the low-temperature transport reflects the competitive coexistence of metallic and Mott-insulating tendencies, with the metallic resistance smoothly connecting to the NC-CDW state, consistent with a supercooled NC-CDW--derived metal \cite{yoshida2015memristive, yoshida2019charge, yin2024real}.
Thus the low temperature resistance ($R_{total}$) can be modeled as a mixture of metallic ($R_{ML}$$\sim {T^2}$) and Mott-like behavior ($R_{MI}$ $\sim e^{\Delta / k_{B}T}$) \cite{ma2016metallic}:
\begin{equation}
    R_{\text{total}} = \phi(\alpha + \beta T^2) + (1 - \phi) e^{\Delta / k_{B}T},
\end{equation} presented in Fig.~S4(b),
where $\phi$ is the ML fraction and $\Delta$ is the Mott gap.
Notably, this metallicity appeared spontaneously, without external perturbation. Previous studies reported that low-temperature metallicity in 1T-TaS$_2$ appears only in flakes thinner than $\sim 31$~nm, attributed to enhanced metastability at reduced thickness~\cite{yoshida2014controlling,yoshida2015memristive, yu2015gate}. In contrast, our measurements on 30 flakes reveal that metallicity also occurs in much thicker samples, up to 60~nm. Statistically, $\sim 5\%$ of flakes with thickness 40--60~nm and $\sim 20\%$ with thickness 5--20~nm are metallic, while $\sim 15\%$ of flakes in the 15--25~nm range remain Mott insulating to the lowest temperatures (see supplementary Information Sec.~III, atomic force microscopy images with height of two such devices is presented in Fig.~S3(a), \cite{Supple}). These results demonstrate that the low-temperature ground state is governed primarily by the out-of-plane stacking of the C-CDW Star-of-David structure rather than by thickness alone. A similar thickness dependence is reported in 2H-TaS$_2$, where the IC-CDW stabilizes at higher temperatures in thin flakes due to enhanced unit-cell lattice instabilities~\cite{rawat2024symmetry}. Overall, while dimensional confinement modifies the competition between electronic and structural phases, in 1T-TaS$_2$ the ground state is predominantly controlled by stacking configuration.
As discussed, spatially resolved studies of 1T-TaS$_2$ reveal heterogeneous domain textures, indicating that transport near the Mott transition proceeds through percolative metallic backbones embedded in an insulating matrix \cite{ma2016metallic, svetin2017three, stahl2020collapse}. Similar behavior has been extensively studied in disordered Hubbard models, where interaction-driven metal--insulator transitions give rise to electronic phase separation, quantum percolation, and inhomogeneous superconductivity, with numerical studies revealing fractal metallic or superconducting clusters that govern transport through sparse connectivity \cite{roy2025metal,wang2025fractionalization, PhysRevB.54.R3756, PhysRevLett.93.126401, grover2010weak}. While such Hubbard-type and disorder-based frameworks successfully capture equilibrium phase coexistence and percolative transport, they are typically restricted to equilibrium or disorder-averaged descriptions and do not incorporate controlled nonequilibrium driving, such as current or voltage pulses\cite{PhysRevB.54.R3756, PhysRevLett.93.126401, grover2010weak}. Consequently, the dynamical reshaping of phase stability under external bias remains outside their scope. Also, previous phenomenological descriptions of charge-density-wave (CDW) systems have largely employed a complex order parameter with well-defined amplitude and phase, successfully capturing equilibrium CDW phases, stiffness, and collective excitations \cite{morozovska2023landau, mcmillan1975landau, moura2024mcmillan, nakatsugawa2020multivalley, domrose2023light}. In contrast, the present work does not focus on the microscopic dynamics of the CDW amplitude or phase, but instead addresses the mesoscale coexistence and competition between two macroscopically distinct electronic states. Therefore, we adopt a phenomenological free-energy framework in which the order parameter is defined as the local volume fraction, directly encoding the spatial distribution of the competing phases.

We considered,
\begin{equation}
    F [\phi]= \phi^2 (1 - \phi)^2
            = \phi^2-2\phi^3+\phi^4,  
\end{equation}
The free energy $F(\phi)$ represents a double-well potential with minima at $\phi=0$ (Mott insulating) and $\phi=1$ (metallic) and a maximum at $\phi=0.5$ [Fig.~1(b)], where the order parameter $\phi=\text{metallic fraction}/(\text{metallic fraction}+\text{Mott fraction})$ is bounded within $0\le\phi\le1$. A positive coefficient of the highest-order term ensures thermodynamic stability and a finite equilibrium value of $\phi$. An applied current induces a transition from the localized Mott state to the delocalized metallic state through Joule heating without breaking global translational symmetry. Because $\phi$ is a non-negative physical fraction, no symmetry requires invariance under $\phi\rightarrow-\phi$; consequently, the free energy is not constrained to satisfy $F(\phi)=F(-\phi)$, and odd-order terms of $\phi$ is allowed in the phenomenological Landau expansion
 \cite{imada1998metal, Chaikin_Lubensky_1995, bar2018kinetic}. 
 Under an applied current bias, we phenomenologically extend the free-energy functional as
\begin{equation}
F(\phi,I)=\phi^{2}(1-\phi)^{2}+a I \phi - b I^{2}\phi^{2},
\end{equation}
where the additional terms represent the leading-order coupling between the electronic state and the external perturbation. In this minimal model, the linear term describes the current-induced stabilization of the insulating configuration in an intermediate bias range, which may arise from enhanced carrier scattering at domain boundaries, increased CDW pinning, and the associated thermal bottleneck preceding collective CDW motion. This scenario is consistent with earlier reports of electric-field-induced CDW stiffening and delayed sliding \cite{mingtao2006negative,lee2018electric,fleming1980electric,bardeen1986depinning}, and is schematically illustrated in Fig.~S5(a). Where for I = 0.5, the Mott state is becoming more favorable, further increasing I to 1.2, take the system towards to metallic state ($I$ is a control parameter (in arbitrary units) representing the perturbation, here analogous to the applied current). The quadratic term represents the effect of Joule heating across domains ($b \propto R t$, with $b>0$), which promotes domain melting and CDW sliding toward the metallic minimum. Within this phenomenological framework, above a threshold current $I_{\mathrm{th}}$ the metallic state becomes energetically favorable, i.e., $F(\phi=1) \leq F(\phi=0)$, yielding $I_{\mathrm{th}} \propto a/b$ [inset of Fig.~1(b)]. For simplicity, a symmetric unperturbed double-well potential is assumed, although in practice the Mott-insulating state is slightly more stable than the metallic state \cite{lee2019origin}. This model provides a compact phenomenological framework that we use in the following to interpret our experimentally observed current-driven evolution between the Mott-insulating and metallic phases.

Figure~2(a) displays the current-induced voltage characteristics of another Mott device. At $T = 10$~K, the initial bias cycles ($n=1-9$) exhibit a nearly linear response, followed by a discrete resistance jump around 2.2~mA, equivalent to free energy modification for current parameter increase up-to $I = 0.5$ (see Fig. S5(a)), where the Mott state becomes more favorable, due to scattering at domain boundary before sliding,  pinned the CDW state \cite{mingtao2006negative, lee2018electric, fleming1980electric, bardeen1986depinning} it corresponds to $a>b$ situation analogous with $a/b = 2$ in Fig.~1(b) inset. From $n = 12$ onward, a progressive decrease in resistance is observed and a negative differential resistance (NDR) region emerges beyond 2.5~mA due to increase value of b ($\propto R t$). By $n = 18$, a pronounced NDR appears beyond 6~mA, culminating in a sharp upturn near 40~mA, indicative of a metallic-like regime (see inset). Due to Joule heating accumulated over cycles, the third term (joule heating) becomes more dominant than the scattering contribution, effectively corresponding to a current parameter $I > 0.5$ analogous with $a/b = 0.5$ (see Fig.~S5(a), Fig.~1(b) inset) \cite{jarach2022joule, li2016joule, PhysRevApplied.10.021001, mihailovic2021ultrafast, yang2024current, mohammadzadeh2021evidence}. As a result, the metallic state becomes more favorable, leading to a sharp drop in resistance accompanied by NDR. By cycle $n = 18$, the increased sweep corresponds to a higher effective $I$, requiring a slightly larger current (6~mA) to induce NDR due to the enhanced scattering term \cite{mingtao2006negative, lee2018electric, fleming1980electric, bardeen1986depinning}. Beyond 40~mA, the nearly linear $V$--$I$ behavior indicates complete dominance of the Joule heating term and a transition of the system toward $\phi = 1$, corresponding to a predominantly metallic, channel-dominated state. For a fixed current value, with increase in cycle $n$, $b$ will keep on increasing ultimately inducing transition towards $\phi = 1$ state, but might require a much larger value of $n$. In contrast, device 1 presented in Fig.~1(a) (indicated in red), subjected to a limited current sweep ($\pm$50~$\mu$A), exhibits no sign of NDR, but the consistent presence of sudden increase in resistance observed within the explored range suggests $I < 0.5$, where scattering term completely dominates ($a>>b$), suggesting requirement of higher joule heating to induce NDR (see Fig.~S6(a), supplementary section V, \cite{Supple}) \cite{mingtao2006negative, lee2018electric, fleming1980electric, bardeen1986depinning}. However, a pristine Mott state ($\phi = 0$) which is deep in the $\phi = 0$ valley, may not evolve into a metallic state unless the current exceeds a high threshold value $I_{th} \propto 1/\phi$, which risks damaging the device, observed in several devices. The behavior of the device in Fig.~2(a) is particularly noteworthy: its Mott-like response along with moderate initial resistance at 10$~K$ ($R = 800\ \Omega$) (see Fig. S5(b))  suggests an initial intermediate order parameter ($\phi \approx 0.5$), enabling a feasible Mott-to-metallic evolution by applying current (Device fabrication is discussed in the supplementary section VIII \cite{Supple}). Overall, for a Joule-heating-driven Mott--metal transition to be feasible, the system must lie near $\phi = 0.5$ in the free-energy landscape.  

Hence, the applied current navigates the system along three distinct trajectories: a regime with \( \frac{dV}{dI} > 0 \) (yellow) corresponds to the fragmentation of domains into coexisting insulating and conductive regions; a regime with \( \frac{dV}{dI} < 0 \) (pink) indicates a percolative dominance of conductive domains, leading to a reduction in overall resistance; and a regime with \( \frac{dV}{dI} = 0 \) (green) marks a dynamically balanced configuration characterized by uniformly distributed resistive phases, presented in Fig.~S6(b) \cite{Supple}. Among the three available path which one system should follow depends on the optimal channel domain configuration which lead to least power dissipation among all admissible paths \cite{Prigogine1947, klein1954principle, guo2017thermodynamic, BertolaCafaro2008}. These regimes reflect current-driven non-equilibrium phase transitions mediated through real-space domain reorganization, revealing the system’s sensitivity to external bias and highlighting the intricate interplay between electronic correlations and structural inhomogeneities~\cite{van2001negative, kumar2017physical, mingtao2006negative} (see supplementary section V \cite{Supple}).

A similar protocol was applied to the low-temperature metal-like (ML) state of device 2 (indicated in black in Fig.~1(a)), with current swept from $-60$~mA to $60$~mA in 1~mA steps (Fig.~2(b)). The black curve corresponding to 10~K temperature, shows a resistance drop from 176~$\Omega$ to 58~$\Omega$, followed by pronounced NDR beyond 12~mA corresponding to transition towards $\phi = 1$ state driven by dominance of joule heating term over the scattering term ($ b >a $), in the very first sweep. Since the low-temperature metallic state implies a reduced population of insulating domains, scattering at CDW boundaries is significantly weaker than in the preceding regime. Rapid CDW domain reorganization (10--100~ps) lowers resistive energy, yielding a threefold drop and robust NDR~\cite{hall1984negative, van2001negative}. Unlike the $R$--$T$ behavior, where the resistance increases with $T$, a current bias drives a steady-state breakdown of residual Mott-insulating domains within the metallic matrix. The system thereby follows a nonequilibrium pathway characterized by $\frac{dV}{dI}<0$, distinct from the equilibrium trajectory. This behavior may be understood in the spirit of the minimum power dissipation principle, which states that a system driven slightly out of equilibrium, can reorganize its internal pathways so as to reduce the overall rate of energy dissipation \cite{Prigogine1947, klein1954principle, guo2017thermodynamic, BertolaCafaro2008}. This problem has been addressed by Onsager and Prigogine, who showed that this holds true for a network of linear resistors with the constraints of current and voltage biasing respectively \cite{Prigogine1947, klein1954principle, guo2017thermodynamic, BertolaCafaro2008}. In our experiments, we encounter a more complex problem where the resistances themselves can vary with the application of a current due to phase transition physics \cite{Prigogine1947, klein1954principle, guo2017thermodynamic, BertolaCafaro2008}. While a concrete proof of the principle in such a case in not known, it is pertinent to check experimentally what are the limits of validity for the principle of least power dissipation in this case. Unlike classical CDW systems (NbSe$_3$, TaS$_3$), NDR here spans the full temperature range and depends sensitively on the initial low-$T$ state. Following~\cite{sherwin1988switching}, NDR emerges when
\[
\frac{dV}{dI} = \frac{1}{\sigma_{CDW} + \sigma_N} - \frac{I \sigma'_{CDW}}{(\sigma_{CDW} + \sigma_N)^2} \leq 0,
\]
which requires \( \sigma'_{CDW} > (\sigma_{CDW} + \sigma_N)/I \), satisfied under sudden domain fragmentation. By the principle of minimum power dissipation, this drives the system into a low-dissipation, NDR phase.
At 100~K, resistance drops from 176~$\Omega$ (10~K) to 78~$\Omega$, then to 28~$\Omega$ $via$ NDR (Fig.~2(b)). Threshold current, $I_{\text{th}}$, increases with $T$ up to 150~K, then sharply decreases above 220~K (Fig.~S7). NDR steps and sharpness reduce with $T$, while threshold power rises linearly till 100~K and drops thereafter (Fig.~2(b), inset). Beyond $I_{\text{th}}$, background MI domains fragment, forming percolative channels responsible for NDR. Pre-formed channels at higher $T$ require higher current/power for fragmentation, explaining the rise in $I_{\text{th}}$ up to 200~K. Upto 150~K the continuous decrese in voltage beyond NDR indicates, $\phi$ corresponding to the system lies in between 0.5 and 1. Above this temperature, the system is more metallic ($\phi \to 1$), needing less drive to disrupt remaining MI patches. As $\sigma_{\text{CDW}}$ varies less with $T$, NDR weakens. At 250~K, voltage rises beyond 53~mA, deviating from the 200~K trend, marking $\phi$ approaching towards 1; at 300~K, full fragmentation yields $dI/dV>0$, indicating a mostly channel-dominated regime ($\phi \sim 1$) (see Fig.~S8(b), Fig.~S8(c), all measurements details are in supplementary section VIII \cite{Supple}). A simpler resistive model is presented in the supplementary section VI to explain the observed NDR in the metallic system. 

For both the MI and ML states, the threshold power ($P_{\text{th}}$) remains in the microwatt range, insufficient to reach the transition temperature from C-CDW to NC-CDW, consistent with the heat balance relation $P = \gamma \Omega (T_e^6 - T_{ph}^6)$, where $\gamma$ is the electron–phonon coupling strength in $\mathrm{W\,K^{-6}\,m^{-3}}$, $\Omega$ is sample volume \cite{PhysRevLett.102.176802, JOUANNE19751047,10.1021/acsnano.9b02870}. For disordered insulators, $\gamma = 1.85~\mathrm{nW\,\mu m^{-3}\,K^{-6}}$; 1T-TaS$_2$ is known to exhibit stronger coupling, giving $\gamma \sim 10^{7}$--$10^{8}~\mathrm{W\,K^{-6}\,m^{-3}}$. Using $\Omega = 40~\mu\mathrm{m} \times 10~\mathrm{nm} \times 40~\mu\mathrm{m}$, the estimated $T_e$ is only $\sim$30~K, far below the NC-CDW transition, but sufficient to trigger thermal runaway driven by local joule heating and domain avalanches (see supplementary section VII \cite{Supple} )\cite{kumar2017physical, das2023physical, adda2022direct, PhysRevApplied.10.021001, chen2009thermal}. In the above discussion, the response of the device was governed by its position in the free-energy landscape. Figure~3(a) shows $V$–$I$ characteristics of two devices with similar low-$T$ Mott insulating states ($R > 1.5$~k$\Omega$; inset), yet differing 50–150~K dips. The red curve ($\phi \sim 0.5$) shows a higher dip and requires 2.5~mA for resistance surge, while the yellow curve ($\phi < 0.5$) requires only 20~$\mu$A but both fails to show NDR. Clearly red curve suggest more metallic fraction in the system than yellow one, as a result required larger critical current (2.5 mA) (I$_c$) (I$_c$ represents the current corresponding to sudden increase in resistance due to dominance of scattering term) than yellow curve, since scattering at CDW domain boundaries is weaker in the metallic state with sparse domains, hence a larger current ($I_c$) is required to trigger the resistance surge, where the scattering contribution becomes dominant. Another way to drive the system towards the metallic fraction limit $\phi = 1$ is through external perturbations such as voltage pulses, which have previously shown local effects; here, however, we observe their macroscopic signature. A 2~V, 100~kHz voltage pulse applied for 60~s to a Mott device results in a sharper resistance peak in the $V$--$I$ curve that shifts towards a higher critical current, indicating an increase in metallic fraction. Further in Fig.~3(b) (inset), increasing current from 3~$\mu$A to 30~$\mu$A deepens the resistance dip in the 50--150~K range, consistent with current-induced enhancement of the metallic volume fraction, i.e., $\phi \to 1$. The observed changes reflect a change in the position of the system along a rugged free energy landscape where local perturbations re-normalize domain configurations.

The melting of Mott insulating domains in 1T-TaS$_2$ is inherently spatiotemporal and inhomogeneous, occurring via nucleation, growth, and coalescence of metallic regions rather than uniform collapse \cite{ma2016metallic, svetin2017three, stahl2020collapse, sun2018hidden, vaskivskyi2015controlling, ishizaka2011femtosecond, wen2019photoinduced, han2015exploration, cho2016nanoscale}. Previous imaging studies have visualized such dynamic domain rearrangements, showing percolating metallic pathways within the insulating matrix \cite{ma2016metallic, svetin2017three, stahl2020collapse, cho2016nanoscale}, while our macroscopic transport measurements reveal stochastic resistive jumps and hysteresis under current or voltage drive, consistent with nonuniform domain dynamics at larger scales. To capture this multiscale behavior, we employ a time-dependent Ginzburg–Landau (TDGL) framework and define $\phi(x,t)$ as the local metallic fraction ($0\le\phi\le1$), which evolves under applied current driving the system out of equilibrium \cite{moura2024mcmillan, morozovska2023landau, mcmillan1975landau}. Its dynamics are governed by, $\frac{\partial \phi(x,t)}{\partial t} = -\Gamma \frac{\delta F}{\delta \phi} + \eta(x,t)$, where $F$ is the free-energy functional, $\Gamma$ is a kinetic coefficient, and $\eta(x,t)$ represents stochastic fluctuations \cite{morozovska2023landau, mcmillan1975landau}. To further account for the nonlinear and stochastic spatiotemporal evolution of domain interfaces, we incorporate a Kardar--Parisi--Zhang (KPZ)–type description \cite{PhysRevLett.131.247101,PhysRevE.103.042102,PhysRevE.105.034104,PhysRevB.8.3423}. Because the current-driven Mott–metal transition involves nucleation and lateral growth of metallic domains within a disordered CDW background, the resulting dynamics naturally falls into the KPZ universality class, independent of microscopic details, providing a unified description of electrically driven domain evolution and macroscopic transport. The KPZ equation gives, $\partial \phi(x,t) / \partial t = \nu \nabla^2 \phi(x,t) + 2\lambda (\nabla \phi(x,t))^2 + \eta (x,t)$, captures the nonlinear growth dynamics of channels in the presence of perturbations, where $\nu$ is a diffusion-like coefficient promoting spatial smoothing, $\lambda$ quantifies the strength of nonlinear coupling between gradients, and $\eta(x,t)$ represents a stochastic Gaussian white noise source. Combining the two yields the generalized evolution equation:  $- m\Gamma \, \delta F / \delta \phi= \nu \nabla^2 \phi(x,t) + 2\lambda (\nabla \phi(x,t)^2 + \eta(x,t) $, where m is  Lagrangian multiplier. Due to application of current, and subsequently setting it zero, the system is initially driven to non equilibrium phase with order parameter having spatiotemporal dependence. Here, the interaction range $p$ between neighboring domains is much smaller than the device dimension $l$ $(p \ll l)$. Consequently, within the device-length scale probed in our measurements, the spatiotemporal evolution of the system can be described using a continuum, coarse-grained framework. Since the measured response averages over a large number of domains, the order parameter defined as the local domain fraction varies smoothly on the relevant length scales \cite{kardar2007statistical, nayak1996possible}. So magnitude of the wave vector of spatial variation has a range of cut off like $k$=2$\pi/l$. In this limiting case, higher order terms of spatial derivatives of order parameters (second term of KPZ equation) is neglected and also neglected the noise term, approximating $\phi$ near zero, higher order of $\phi$ term has been neglected, then the combined equation for our case deduce to $\nu \nabla^2 \phi(x) = -m\Gamma \, \delta F / \delta \phi$. From F = $\phi^2(1-\phi)^2$, $\nu \nabla^2 \phi(x) = -2m\Gamma \phi$. In 1D positive value of m gives periodic solution of $\phi$, which is nonphysical in this stochastic dynamics. m = 0 gives trivial solution, hence considering  negative value of m, and the bound nature of $\phi$, $\phi$ normalized to $e^{-x/d}$. Where $d = \sqrt{2m\Gamma/\nu}$ is the characteristic length scale of channel and $x/d$ represents domain to channel ratio. As $d \to 0$, the order parameter saturates to $\phi \to 0$, signifying a state dominated by domains i.e the Mott state, $d \to \infty$, the order parameter saturates to $\phi \to 1$, indicating channel dominated metallic state. Fig. 4(a) shows for small value of domain to channel ratio, $\phi$ easily achieves 1 and for larger value,  $\phi$ goes sway from 1, which well explain our observations. 
The KPZ framework predicts universal scaling of domain-wall fluctuations characterized by roughness and growth exponents \cite{PhysRevLett.131.247101,PhysRevE.103.042102,PhysRevE.105.034104,PhysRevB.8.3423}. A direct extraction of these exponents requires systematic variation of the interfacial length scale, the present work focuses on a phenomenological description of current-driven domain dynamics, while future measurements on devices with tunable channel lengths or real-space probes will enable a direct test of KPZ universality and a broader generalization of the model. Also, further experiments combining transport with in-operando structural probes will enable a more direct visualization of the domain dynamics described here.
 
The non-linear conductivity in 1T-TaS\(_2\) follows a power-law, \( \sigma \propto (I - I_{\text{th}})^\beta \), analogous to the percolation scaling \( \sigma \propto (p - p_c)^t \), where \(p\) is the fraction of conducting sites, \(p_c\) the percolation threshold, and \(t \approx 1.3\) is the universal 2D exponent~\cite{stauffer, berg2023percolation}. For \(T \leq 100~\text{K}\), \(\beta \approx 1.3\) confirms a percolative transport regime. At 150 K, \(\beta\) peaks at 1.48, coinciding with maxima in \(I_{\text{th}}\) and \(P_{\text{th}}\), where domain fragmentation dominates and sharp conductivity jumps occur. Above 150 K, \(\beta\) decreases to \(\approx 0.8\) at 300 K, indicating a transition to a channel-driven regime with smoother I-V characteristics (Fig.~2(b)). In the sparse fractal limit, the system exhibits enhanced non-linearity due to filamentary percolation pathways that dynamically evolve under current bias. The fractal nature of the system can be realized by analyzing the area across the NDR when the system transition from higher resistive state to more metallic channel dominated lower resistive state. From statistical mechanics, the number of active conduction channels within these regions scales with characteristic linear size of these metallic patches as $N \sim L^{D_f}$, where $D_f$ is the fractal dimension \cite{malarz2025confirming}. During a current-driven transition, for sample lengths \(L \gg \xi\), the increase in current above the threshold, $|I-I_\mathrm{th}|$, activates additional metallic clusters, so that $|I-I_\mathrm{th}| \sim N \sim L^{D_f}$ and thus $L \sim |I-I_\mathrm{th}|^{1/D_f}$. The area of NDR reflects substantial domain reconfigurations during the transition from insulating to metallic states, with the fractal dimension governing how these reconfigurations scale with the applied current, which gives, $A_\mathrm{NDR} \sim |I-I_\mathrm{th}|^{2/D_f}$. At low temperatures, $D_f\sim0.37$ indicates that metallic regions are sparse and filamentary, while at 100 K, $D_f\sim0.5$ suggests more connected, yet still inhomogeneous domains. As temperature increases further, $D_f$ approaches $\sim0.9$ at 300 K, consistent with nearly uniform metallic regions.  As \(D_f\) increases, denser pathways form and the switching time $t_{switch}$ decreases, highlighting the geometric control of transport kinetics. In the context of fractality, post-NDR oscillations follow conductance \(G \propto (I_p - I_{th})^\delta\), where I$_p$ is current corresponding to each oscillation post-NDR, with \(\delta = 0.3\) for Mott state, (Fig.~2(a)), 0.4–0.5 for metallic state (Fig.~2(b)) presented in Fig.~S11, reflecting a multiscale percolation hierarchy that governs non-linear conduction $via$ evolving fractal connectivity in the sparse-limit ~\cite{stauffer}. 
Overall, our work provides a complementary macroscopic perspective on domain dynamics in 1T-TaS$_2$. Whereas prior microscopic studies, such as STM and STS, have predominantly focused on imaging domain evolution under external perturbations at length scales of 20--500~nm, our transport measurements, combined with a phenomenological free-energy framework and KPZ dynamics, capture the collective nonequilibrium evolution of domains at the device scale \cite{ma2016metallic, svetin2017three, stahl2020collapse, cho2016nanoscale}. In this regime, the volume fraction of domains evolves as $e^{-x/d}$ under external perturbations, where $\frac{x}{d}$ represents the domain-to-channel ratio. Our results are consistent with earlier observations, while further analysis demonstrates that domain reconfiguration driven by external stimuli is governed by a fractal dimension $D_f$, leading naturally to percolative transport behavior.

Inspired by percolation models~\cite{stauffer}, we construct a current-tuned site-percolation framework to explain the evolution of non-linear transport in 1T-TaS\(_2\). A 2D block of \(n \times n\) sites is modeled with open-site probability \(p\), initially increasing linearly with current \(I\). Below the percolation threshold \(p_c = 0.59\), the system exhibits linear I-V behavior. Beyond a threshold current, non-linearity emerges, and the transition is captured by a modified control function \( (p_c - p)/(1 - rp) \), with \(r\) a coupling parameter, analogous to \( (I_{th} - I)/(1 - rI) \), provided \(rI < 1\). Expansion yields a dominant quadratic term \( \sim I^2 \), signifying local Joule heating, which activates conducting sites and triggers NDR near \(I_{th}\) (Fig.~4(b)). This model connects the emergence of NDR to a thermally driven transition from insulating (\(\phi=0\)) to metallic (\(\phi=1\)) states. Whereas, in the Mott phase, considering site opening is $\propto$ current, shows absence of NDR in V-I curve presented in Fig.S10(a), suggests insufficient Joule heating prevents transition towards $\phi$ = 1, thus suppressing NDR. The sparse fractal regime amplifies these effects due to enhanced local field concentration and filamentary conduction paths, reinforcing the observed scaling behavior and non-linear dynamics~\cite{berg2023percolation}. Our analysis provides a minimal yet physically consistent picture of current-induced percolative switching and NDR emergence in correlated cdw systems (see supplementary section IX, Fig.~S10, \cite{Supple}).

\section*{Conclusion}
In conclusion, our study of MI and ML states in 1T-TaS\(_2\) under current bias combines transport experiments with a minimal free-energy model \(F[\phi,I]\), capturing threshold-driven stabilization of metallic domains and easier transitions at high bias. Pulse-driven measurements reveal intrinsic switching, with the NDR onset exhibiting a scaling exponent consistent with 2D percolation (\(\beta \approx 1.3\)). The sparse limit fractal dimension \(D_f \approx 0.3\) at 10~K, rising to \(\approx 0.9\) at 300~K, indicates temperature-driven multiscale pathway development. In the MI state, domain stability up to 190~K yields nearly constant $I_{th}$, while above 220~K enhanced conduction marks the emergence of a channel-dominated IC-CDW phase. In the ML state, an increase in temperature demands greater current and power to fragment domains before forming continuous pathways, elucidating the temperature-dependent interaction of percolation, domain reconfiguration, and least power dissipation. These results establish a quantitative link between current-driven phase transitions, fractal geometry, and percolative transport, offering a general framework for engineered domain architectures in low-dimensional correlated electron systems.

\section*{acknowledgement}
The financial support from UGC (Certificate No: JUN21C07016), and IACS are greatly acknowledged. PM thanks all the CSS operators of IACS. SD acknowledges the financial support from DST-SERB grant No. CRG/2021/004334. SD also acknowledges support from the Technical Research Centre (TRC), IACS, Kolkata.

Conceptualization and resources were provided by S.D. Crystal growth was performed by K.R. Transport (dc and pulse) measurements and data curation were carried out at IACS by P.M., supported by K.D., B.D. Electron localization calculation was done by P.M and T.K. The second harmonic response was measured at IACS by P.M., supported by B.D. Device fabrication was conducted at IACS by M.P., P.M., S.M. with technical support provided by P.K.H., B.K. Resistive network analysis was performed by B.M. in consultation with S.D. and P.M. The computational approach was led by S.G., supported by M.A.M. XPS measurements were performed by S.K.M. and S.C., supported by K.R. The power dissipation model was proposed by S.S. The free-energy model and related calculations were developed by S.D. and P.M. P.M is thankful to Soham Das for useful discussion. The manuscript was written by P.M. and S.D., with input from all authors.

\bibliography{ref}

@Preamble
{
	"\providecommand{\noopsort}[1]{}"
	# "\providecommand{\singleletter}[1]{#1}%"
}

@article{PhysRevLett.125.097002,
  title = {Enhanced Electron-Phonon Coupling for Charge-Density-Wave Formation in ${\mathrm{La}}_{1.8\ensuremath{-}x}{\mathrm{Eu}}_{0.2}{\mathrm{Sr}}_{x}{\mathrm{CuO}}_{4+\ensuremath{\delta}}$},
  author = {Peng, Y. Y. and Husain, A. A. and Mitrano, M. and Sun, S. X.-L. and Johnson, T. A. and Zakrzewski, A. V. and MacDougall, G. J. and Barbour, A. and Jarrige, I. and Bisogni, V. and Abbamonte, P.},
  journal = {Phys. Rev. Lett.},
  volume = {125},
  issue = {9},
  pages = {097002},
  numpages = {6},
  year = {2020},
  month = {Aug},
  publisher = {American Physical Society},
  
}

@article{PhysRevB.107.024101,
  title = {Evolution of static charge density wave order, amplitude mode dynamics, and suppression of Kohn anomalies at the hysteretic transition in ${\mathrm{EuTe}}_{4}$},
  author = {Rathore, Ranjana and Pathak, Abhishek and Gupta, Mayanak K. and Mittal, Ranjan and Kulkarni, Ruta and Thamizhavel, A. and Singhal, Himanshu and Said, Ayman H. and Bansal, Dipanshu},
  journal = {Phys. Rev. B},
  volume = {107},
  issue = {2},
  pages = {024101},
  numpages = {9},
  year = {2023},
  month = {Jan},
  publisher = {American Physical Society},
  
}

@article{zhu2015classification,
  title={Classification of charge density waves based on their nature},
  author={Zhu, Xuetao and Cao, Yanwei and Zhang, Jiandi and Plummer, EW and Guo, Jiandong},
  journal={Proceedings of the National Academy of Sciences},
  volume={112},
  number={8},
  pages={2367--2371},
  year={2015},
  publisher={National Acad Sciences}
}

@article{PhysRev.131.1995,
  title = {Theory of Kohn Anomalies in the Phonon Spectra of Metals},
  author = {Taylor, P. L.},
  journal = {Phys. Rev.},
  volume = {131},
  issue = {5},
  pages = {1995--1999},
  numpages = {0},
  year = {1963},
  month = {Sep},
  publisher = {American Physical Society},
  
}

@article{PhysRevB.3.3173,
  title = {Observability of Charge-Density Waves by Neutron Diffraction},
  author = {Overhauser, A. W.},
  journal = {Phys. Rev. B},
  volume = {3},
  issue = {10},
  pages = {3173--3182},
  numpages = {0},
  year = {1971},
  month = {May},
  publisher = {American Physical Society},
  
}

@article{PhysRev.167.691,
  title = {Exchange and Correlation Instabilities of Simple Metals},
  author = {Overhauser, A. W.},
  journal = {Phys. Rev.},
  volume = {167},
  issue = {3},
  pages = {691--698},
  numpages = {0},
  year = {1968},
  month = {Mar},
  publisher = {American Physical Society},
  
}

@article{friend1979periodic,
  title={Periodic lattice distortions and charge density waves in one-and two-dimensional metals},
  author={Friend, R Ho and Jerome, D},
  journal={Journal of Physics C: Solid State Physics},
  volume={12},
  number={8},
  pages={1441},
  year={1979},
  publisher={IOP Publishing}
}

@article{rice1973theory,
  title={Theory of a quasi-one-dimensional band-conductor},
  author={Rice, MJ and Str{\"a}ssler, S},
  journal={Solid State Communications},
  volume={13},
  number={1},
  pages={125--128},
  year={1973},
  publisher={Elsevier}
}

@article{PhysRevB.77.165135,
  title = {Fermi surface nesting and the origin of charge density waves in metals},
  author = {Johannes, M. D. and Mazin, I. I.},
  journal = {Phys. Rev. B},
  volume = {77},
  issue = {16},
  pages = {165135},
  numpages = {8},
  year = {2008},
  month = {Apr},
  publisher = {American Physical Society},
  
}

@article{hoesch2009giant,
  title={Giant Kohn anomaly and the phase transition in charge density wave ZrTe 3},
  author={Hoesch, Moritz and Bosak, Alexey and Chernyshov, Dmitry and Berger, Helmuth and Krisch, Michael},
  journal={Physical review letters},
  volume={102},
  number={8},
  pages={086402},
  year={2009},
  publisher={APS}
}

@article{rossnagel2011origin,
  title={On the origin of charge-density waves in select layered transition-metal dichalcogenides},
  author={Rossnagel, K},
  journal={Journal of Physics: Condensed Matter},
  volume={23},
  number={21},
  pages={213001},
  year={2011},
  publisher={IOP Publishing}
}

@article{aebi2001search,
  title={On the search for Fermi surface nesting in quasi-2D materials},
  author={Aebi, Philipp and Pillo, Th and Berger, Helmuth and L{\'e}vy, F},
  journal={Journal of electron spectroscopy and related phenomena},
  volume={117},
  pages={433--449},
  year={2001},
  publisher={Elsevier}
}

@article{shen2008primary,
  title={Primary role of the barely occupied states in the charge density wave formation of NbSe 2},
  author={Shen, DW and Zhang, Y and Yang, LX and Wei, J and Ou, HW and Dong, JK and Xie, BP and He, C and Zhao, JF and Zhou, B and others},
  journal={Physical review letters},
  volume={101},
  number={22},
  pages={226406},
  year={2008},
  publisher={APS}
}

@article{bardeen1980tunneling,
  title={Tunneling theory of charge-density-wave depinning},
  author={Bardeen, John},
  journal={Physical Review Letters},
  volume={45},
  number={24},
  pages={1978},
  year={1980},
  publisher={APS}
}

@article{hall1984negative,
  title={Negative Differential Resistance and Instability in Nb Se 3},
  author={Hall, RP and Sherwin, M and Zettl, A},
  journal={Physical review letters},
  volume={52},
  number={25},
  pages={2293},
  year={1984},
  publisher={APS}
}

@article{PhysRevLett.87.126401,
  title = {Negative Resistance and Local Charge-Density-Wave Dynamics},
  author = {van der Zant, H. S. J. and Slot, E. and Zaitsev-Zotov, S. V. and Artemenko, S. N.},
  journal = {Phys. Rev. Lett.},
  volume = {87},
  issue = {12},
  pages = {126401},
  numpages = {4},
  year = {2001},
  month = {Aug},
  publisher = {American Physical Society},
  
}

@article{tsen2015structure,
  title={Structure and control of charge density waves in two-dimensional 1T-TaS2},
  author={Tsen, Adam W and Hovden, Robert and Wang, Dennis and Kim, Young Duck and Okamoto, Junichi and Spoth, Katherine A and Liu, Yu and Lu, Wenjian and Sun, Yuping and Hone, James C and others},
  journal={Proceedings of the National Academy of Sciences},
  volume={112},
  number={49},
  pages={15054--15059},
  year={2015},
  publisher={National Acad Sciences}
}

@article{lee2019origin,
  title={Origin of the insulating phase and first-order metal-insulator transition in 1 t-tas 2},
  author={Lee, Sung-Hoon and Goh, Jung Suk and Cho, Doohee},
  journal={Physical review letters},
  volume={122},
  number={10},
  pages={106404},
  year={2019},
  publisher={APS}
}

@article{yoshida2014controlling,
  title={Controlling charge-density-wave states in nano-thick crystals of 1T-TaS2},
  author={Yoshida, Masaro and Zhang, Yijin and Ye, Jianting and Suzuki, Ryuji and Imai, Yasuhiko and Kimura, Shigeru and Fujiwara, Akihiko and Iwasa, Yoshihiro},
  journal={Scientific reports},
  volume={4},
  number={1},
  pages={7302},
  year={2014},
  publisher={Nature Publishing Group UK London}
}

@article{Rabi,
  title={Percolation and tortuosity in heart-like cells},
  author={R. Rabinovitch and Y. Biton and D. Braunstein and I.Aviram and R. Thieberger and A. Rabinovitch},
  journal={Scientific Reports},
  volume={11},
  pages={11441},
  year={2021},
  publisher={Springer Nature}
}

@book{stauffer,
  title={Introduction to Percolation Theory},
  author={D. Stauffer and A. Aharony},
  edition={2},
  year={1992},
  publisher={Taylor \& Francis}
}

@article{wu2022effect,
  title={Effect of stacking order on the electronic state of 1 T-TaS 2},
  author={Wu, Zongxiu and Bu, Kunliang and Zhang, Wenhao and Fei, Ying and Zheng, Yuan and Gao, Jingjing and Luo, Xuan and Liu, Zheng and Sun, Yu-Ping and Yin, Yi},
  journal={Physical Review B},
  volume={105},
  number={3},
  pages={035109},
  year={2022},
  publisher={APS}
}

@article{wang2019lattice,
  title={Lattice discontinuities of 1T-TaS2 across first order charge density wave phase transitions},
  author={Wang, Wen and Dietzel, Dirk and Schirmeisen, Andr{\'e}},
  journal={Scientific reports},
  volume={9},
  number={1},
  pages={7066},
  year={2019},
  publisher={Nature Publishing Group UK London}
}

@article{stahl2020collapse,
  title={Collapse of layer dimerization in the photo-induced hidden state of 1T-TaS2},
  author={Stahl, Quirin and Kusch, Maximilian and Heinsch, Florian and Garbarino, Gaston and Kretzschmar, Norman and Hanff, Kerstin and Rossnagel, Kai and Geck, Jochen and Ritschel, Tobias},
  journal={Nature communications},
  volume={11},
  number={1},
  pages={1247},
  year={2020},
  publisher={Nature Publishing Group UK London}
}

@article{ma2016metallic,
  title={A metallic mosaic phase and the origin of Mott-insulating state in 1T-TaS2},
  author={Ma, Liguo and Ye, Cun and Yu, Yijun and Lu, Xiu Fang and Niu, Xiaohai and Kim, Sejoong and Feng, Donglai and Tom{\'a}nek, David and Son, Young-Woo and Chen, Xian Hui and others},
  journal={Nature communications},
  volume={7},
  number={1},
  pages={10956},
  year={2016},
  publisher={Nature Publishing Group UK London}
}

@article{yoshida2015memristive,
  title={Memristive phase switching in two-dimensional 1T-TaS2 crystals},
  author={Yoshida, Masaro and Suzuki, Ryuji and Zhang, Yijin and Nakano, Masaki and Iwasa, Yoshihiro},
  journal={Science advances},
  volume={1},
  number={9},
  pages={e1500606},
  year={2015},
  publisher={American Association for the Advancement of Science}
}

@article{sun2018hidden,
  title={Hidden CDW states and insulator-to-metal transition after a pulsed femtosecond laser excitation in layered chalcogenide 1T-TaS2- x Se x},
  author={Sun, Kai and Sun, Shuaishuai and Zhu, Chunhui and Tian, Huanfang and Yang, Huaixin and Li, Jianqi},
  journal={Science advances},
  volume={4},
  number={7},
  pages={eaas9660},
  year={2018},
  publisher={American Association for the Advancement of Science}
}

@article{vaskivskyi2015controlling,
  title={Controlling the metal-to-insulator relaxation of the metastable hidden quantum state in 1T-TaS2},
  author={Vaskivskyi, Igor and Gospodaric, Jan and Brazovskii, Serguei and Svetin, Damjan and Sutar, Petra and Goreshnik, Evgeny and Mihailovic, Ian A and Mertelj, Tomaz and Mihailovic, Dragan},
  journal={Science advances},
  volume={1},
  number={6},
  pages={e1500168},
  year={2015},
  publisher={American Association for the Advancement of Science}
}

@article{ishizaka2011femtosecond,
  title={Femtosecond core-level photoemision spectroscopy on 1 T-TaS 2 using a 60-eV laser source},
  author={Ishizaka, K and Kiss, T and Yamamoto, T and Ishida, Y and Saitoh, T and Matsunami, M and Eguchi, R and Ohtsuki, T and Kosuge, A and Kanai, T and others},
  journal={Physical Review B—Condensed Matter and Materials Physics},
  volume={83},
  number={8},
  pages={081104},
  year={2011},
  publisher={APS}
}

@article{wen2019photoinduced,
  title={Photoinduced phase transitions in two-dimensional charge-density-wave 1T-TaS2},
  author={Wen, Wen and Dang, Chunhe and Xie, Liming},
  journal={Chinese Physics B},
  volume={28},
  number={5},
  pages={058504},
  year={2019},
  publisher={IOP Publishing}
}

@article{han2015exploration,
  title={Exploration of metastability and hidden phases in correlated electron crystals visualized by femtosecond optical doping and electron crystallography},
  author={Han, Tzong-Ru T and Zhou, Faran and Malliakas, Christos D and Duxbury, Phillip M and Mahanti, Subhendra D and Kanatzidis, Mercouri G and Ruan, Chong-Yu},
  journal={Science advances},
  volume={1},
  number={5},
  pages={e1400173},
  year={2015},
  publisher={American Association for the Advancement of Science}
}

@article{hollander2015electrically,
  title={Electrically driven reversible insulator--metal phase transition in 1T-TaS2},
  author={Hollander, Matthew J and Liu, Yu and Lu, Wen-Jian and Li, Li-Jun and Sun, Yu-Ping and Robinson, Joshua A and Datta, Suman},
  journal={Nano letters},
  volume={15},
  number={3},
  pages={1861--1866},
  year={2015},
  publisher={ACS Publications}
}

@article{yu2015gate,
  title={Gate-tunable phase transitions in thin flakes of 1T-TaS2},
  author={Yu, Yijun and Yang, Fangyuan and Lu, Xiu Fang and Yan, Ya Jun and Cho, Yong-Heum and Ma, Liguo and Niu, Xiaohai and Kim, Sejoong and Son, Young-Woo and Feng, Donglai and others},
  journal={Nature nanotechnology},
  volume={10},
  number={3},
  pages={270--276},
  year={2015},
  publisher={Nature Publishing Group UK London}
}

@article{van2001negative,
  title={Negative resistance and local charge-density-wave dynamics},
  author={Van der Zant, HSJ and Slot, E and Zaitsev-Zotov, SV and Artemenko, SN},
  journal={Physical Review Letters},
  volume={87},
  number={12},
  pages={126401},
  year={2001},
  publisher={APS}
}

@article{wang2024dualistic,
  title={Dualistic insulator states in 1T-TaS2 crystals},
  author={Wang, Yihao and Li, Zhihao and Luo, Xuan and Gao, Jingjing and Han, Yuyan and Jiang, Jialiang and Tang, Jin and Ju, Huanxin and Li, Tongrui and Lv, Run and others},
  journal={Nature Communications},
  volume={15},
  number={1},
  pages={3425},
  year={2024},
  publisher={Nature Publishing Group UK London}
}

@article{petkov2022atomic,
  title={Atomic structure and Mott nature of the insulating charge density wave phase of 1T-TaS2},
  author={Petkov, Valeri and Peralta, Juan E and Aoun, B and Ren, Y},
  journal={Journal of Physics: Condensed Matter},
  volume={34},
  number={34},
  pages={345401},
  year={2022},
  publisher={IOP Publishing}
}

@article{dardel1992spectroscopic,
  title={Spectroscopic signatures of phase transitions in a charge-density-wave system: 1T-TaS 2},
  author={Dardel, B and Grioni, M and Malterre, D and Weibel, P and Baer, Y and L{\'e}vy, F},
  journal={Physical Review B},
  volume={46},
  number={12},
  pages={7407},
  year={1992},
  publisher={APS}
}

@article{sherwin1988switching,
  title={Switching and charge-density-wave transport in NbSe 3. III. Dynamical instabilities},
  author={Sherwin, MS and Zettl, A and Hall, RP},
  journal={Physical Review B},
  volume={38},
  number={18},
  pages={13028},
  year={1988},
  publisher={APS}
}

@article{kumar2017physical,
  title={Physical origins of current and temperature controlled negative differential resistances in NbO2},
  author={Kumar, Suhas and Wang, Ziwen and Davila, Noraica and Kumari, Niru and Norris, Kate J and Huang, Xiaopeng and Strachan, John Paul and Vine, David and Kilcoyne, AL David and Nishi, Yoshio and others},
  journal={Nature communications},
  volume={8},
  number={1},
  pages={658},
  year={2017},
  publisher={Nature Publishing Group UK London}
}

@article{das2023physical,
  title={Physical Origin of Negative Differential Resistance in V3O5 and Its Application as a Solid-State Oscillator},
  author={Das, Sujan Kumar and Nandi, Sanjoy Kumar and Marquez, Camilo Verbel and R{\'u}a, Armando and Uenuma, Mutsunori and Puyoo, Etienne and Nath, Shimul Kanti and Albertini, David and Baboux, Nicolas and Lu, Teng and others},
  journal={Advanced Materials},
  volume={35},
  number={8},
  pages={2208477},
  year={2023},
  publisher={Wiley Online Library}
}

@article{adda2022direct,
  title={Direct observation of the electrically triggered insulator-metal transition in V 3 O 5 far below the transition temperature},
  author={Adda, Coline and Lee, Min-Han and Kalcheim, Yoav and Salev, Pavel and Rocco, Rodolfo and Vargas, Nicolas M and Ghazikhanian, Nareg and Li, Chung-Pang and Albright, Grant and Rozenberg, Marcelo and others},
  journal={Physical Review X},
  volume={12},
  number={1},
  pages={011025},
  year={2022},
  publisher={APS}
}

@article{mohammadzadeh2021room,
  title={Room temperature depinning of the charge-density waves in quasi-two-dimensional 1T-TaS2 devices},
  author={Mohammadzadeh, Amirmahdi and Rehman, Adil and Kargar, Fariborz and Rumyantsev, Sergei and Smulko, Janusz M and Knap, Wojciech and Lake, Roger K and Balandin, Alexander A},
  journal={Applied Physics Letters},
  volume={118},
  number={22},
  year={2021},
  publisher={AIP Publishing}
}

@phdthesis{mingtao2006negative,
  title={Negative Resistance and Charge-Density-Wave transport},
  author={Mingtao, Lu},
  year={2006}
}

@article{berg2023percolation,
  title={Percolation and conductivity in evolving disordered media},
  author={Berg, Carl Fredrik and Sahimi, Muhammad},
  journal={Physical Review E},
  volume={108},
  number={2},
  pages={024132},
  year={2023},
  publisher={APS}
}

@article{PhysRevB.35.197,
  title = {Transport properties of continuum systems near the percolation threshold},
  author = {Feng, Shechao and Halperin, B. I. and Sen, P. N.},
  journal = {Phys. Rev. B},
  volume = {35},
  issue = {1},
  pages = {197--214},
  numpages = {0},
  year = {1987},
  month = {Jan},
  publisher = {American Physical Society},
  
}

@article{PhysRevB.55.8038,
  title = {Archie's law from a fractal model for porous rocks},
  author = {Roy, Shashwati and Tarafdar, S.},
  journal = {Phys. Rev. B},
  volume = {55},
  issue = {13},
  pages = {8038--8041},
  numpages = {0},
  year = {1997},
  month = {Apr},
  publisher = {American Physical Society},
 
}

@article{PhysRevLett.120.166401,
  title = {Ultrafast Doublon Dynamics in Photoexcited $1T$-${\mathrm{TaS}}_{2}$},
  author = {Ligges, M. and Avigo, I. and Gole\ifmmode \check{z}\else \v{z}\fi{}, D. and Strand, H. U. R. and Beyazit, Y. and Hanff, K. and Diekmann, F. and Stojchevska, L. and Kall\"ane, M. and Zhou, P. and Rossnagel, K. and Eckstein, M. and Werner, P. and Bovensiepen, U.},
  journal = {Phys. Rev. Lett.},
  volume = {120},
  issue = {16},
  pages = {166401},
  numpages = {5},
  year = {2018},
  month = {Apr},
  publisher = {American Physical Society},
  
}

@article{PhysRevLett.131.247101,
  title = {Unpredicted Scaling of the One-Dimensional Kardar-Parisi-Zhang Equation},
  author = {Fontaine, C\^ome and Vercesi, Francesco and Brachet, Marc and Canet, L\'eonie},
  journal = {Phys. Rev. Lett.},
  volume = {131},
  issue = {24},
  pages = {247101},
  numpages = {6},
  year = {2023},
  month = {Dec},
  publisher = {American Physical Society},
  
}

@article{PhysRevE.103.042102,
  title = {Conserved Kardar-Parisi-Zhang equation: Role of quenched disorder in determining universality},
  author = {Mukherjee, Sudip},
  journal = {Phys. Rev. E},
  volume = {103},
  issue = {4},
  pages = {042102},
  numpages = {10},
  year = {2021},
  month = {Apr},
  publisher = {American Physical Society},
 
}

@article{PhysRevE.105.034104,
  title = {Disorders can induce continuously varying universal scaling in driven systems},
  author = {Haldar, Astik and Basu, Abhik},
  journal = {Phys. Rev. E},
  volume = {105},
  issue = {3},
  pages = {034104},
  numpages = {18},
  year = {2022},
  month = {Mar},
  publisher = {American Physical Society},
 
}

@article{PhysRevLett.102.176802,
  title = {Electron-Phonon Decoupling in Disordered Insulators},
  author = {Ovadia, M. and Sac\'ep\'e, B. and Shahar, D.},
  journal = {Phys. Rev. Lett.},
  volume = {102},
  issue = {17},
  pages = {176802},
  numpages = {4},
  year = {2009},
  month = {Apr},
  publisher = {American Physical Society},
  
}

@article{JOUANNE19751047,
title = {Electron-phonon coupling in highly doped n type silicon},
journal = {Solid State Communications},
volume = {16},
number = {8},
pages = {1047-1049},
year = {1975},
issn = {0038-1098},

author = {M. Jouanne and R. Beserman and I. Ipatova and A. Subashiev},
}

@article{10.1021/acsnano.9b02870,
author = {Geremew, Adane
K. and Rumyantsev, Sergey and Kargar, Fariborz and Debnath, Bishwajit and Nosek, Adrian and Bloodgood, Matthew A. and Bockrath, Marc and Salguero, Tina T. and Lake, Roger K. and Balandin, Alexander A.},
title = {Bias-Voltage Driven Switching of the Charge-Density-Wave and Normal Metallic Phases in 1T-TaS2 Thin-Film Devices},
journal = {ACS Nano},
volume = {13},
number = {6},
pages = {7231-7240},
year = {2019},

}

@article{PhysRevB.100.155407,
  title = {Surface structure and stacking of the commensurate $(\sqrt{13}\ifmmode\times\else\texttimes\fi{}\sqrt{13})R13.{9}^{\ensuremath{\circ}}$ charge density wave phase of $1T\ensuremath{-}{\mathrm{TaS}}_{2}(0001)$},
  author = {von Witte, Gevin and Ki\ss{}linger, Tilman and Horstmann, Jan Gerrit and Rossnagel, Kai and Schneider, M. Alexander and Ropers, Claus and Hammer, Lutz},
  journal = {Phys. Rev. B},
  volume = {100},
  issue = {15},
  pages = {155407},
  numpages = {11},
  year = {2019},
  month = {Oct},
  publisher = {American Physical Society},
 
}

@article{ritschel2015orbital,
  title={Orbital textures and charge density waves in transition metal dichalcogenides},
  author={Ritschel, T and Trinckauf, J and Koepernik, K and B{\"u}chner, B and Zimmermann, M v and Berger, H and Joe, YI and Abbamonte, P and Geck, J},
  journal={Nature physics},
  volume={11},
  number={4},
  pages={328--331},
  year={2015},
  publisher={Nature Publishing Group UK London}
}

@article{PhysRevB.8.3423,
  title = {Time-Dependent Ginzburg-Landau Theory of Nonequilibrium Relaxation},
  author = {Binder, K.},
  journal = {Phys. Rev. B},
  volume = {8},
  issue = {7},
  pages = {3423--3438},
  numpages = {0},
  year = {1973},
  month = {Oct},
  publisher = {American Physical Society},
  
}

@article{PhysRevApplied.10.021001,
  title = {Radiative Thermal Runaway Due to Negative-Differential Thermal Emission Across a Solid-Solid Phase Transition},
  author = {Bierman, David M. and Lenert, Andrej and Kats, Mikhail A. and Zhou, You and Zhang, Shuyan and De La Ossa, Matthew and Ramanathan, Shriram and Capasso, Federico and Wang, Evelyn N.},
  journal = {Phys. Rev. Appl.},
  volume = {10},
  issue = {2},
  pages = {021001},
  numpages = {6},
  year = {2018},
  month = {Aug},
  publisher = {American Physical Society},
  
}

@article{vaskivskyi2016fast,
  title={Fast electronic resistance switching involving hidden charge density wave states},
  author={Vaskivskyi, I and Mihailovic, IA and Brazovskii, S and Gospodaric, J and Mertelj, T and Svetin, D and Sutar, P and Mihailovic, D},
  journal={Nature communications},
  volume={7},
  number={1},
  pages={11442},
  year={2016},
  publisher={Nature Publishing Group UK London}
}

@incollection{lee2018electric,
  title={Electric field depinning of charge density waves},
  author={Lee, PA and Rice, TM},
  booktitle={Basic Notions Of Condensed Matter Physics},
  pages={460--470},
  year={2018},
  publisher={CRC Press}
}

@article{fleming1980electric,
  title={Electric-field depinning of charge-density waves in Nb Se 3},
  author={Fleming, RM},
  journal={Physical Review B},
  volume={22},
  number={12},
  pages={5606},
  year={1980},
  publisher={APS}
}

@inproceedings{bardeen1986depinning,
  title={Depinning of charge-density waves by quantum tunneling},
  author={Bardeen, John},
  booktitle={Proceedings of the Yamada Conference XV on Physics and Chemistry of Quasi One-Dimensional Conductors},
  pages={14--18},
  year={1986},
  organization={Elsevier}
}

@article{jarach2022joule,
  title={Joule-heating induced phase transition in 1T-TaS2 near room temperature probed by thermal imaging of power dissipation},
  author={Jarach, Yaron and Rodes, Lior and Ber, Emanuel and Yalon, Eilam and Kanigel, Amit},
  journal={Applied Physics Letters},
  volume={120},
  number={8},
  year={2022},
  publisher={AIP Publishing}
}

@article{li2016joule,
  title={Joule heating-induced metal--insulator transition in epitaxial VO2/TiO2 devices},
  author={Li, Dasheng and Sharma, Abhishek A and Gala, Darshil K and Shukla, Nikhil and Paik, Hanjong and Datta, Suman and Schlom, Darrell G and Bain, James A and Skowronski, Marek},
  journal={ACS applied materials \& interfaces},
  volume={8},
  number={20},
  pages={12908--12914},
  year={2016},
  publisher={ACS Publications}
}

@article{mihailovic2021ultrafast,
  title={Ultrafast non-thermal and thermal switching in charge configuration memory devices based on 1T-TaS2},
  author={Mihailovic, D and Svetin, D and Vaskivskyi, I and Venturini, R and Lipov{\v{s}}ek, B and Mraz, A},
  journal={Applied Physics Letters},
  volume={119},
  number={1},
  year={2021},
  publisher={AIP Publishing}
}

@article{yang2024current,
  title={Current-Driven to Thermally Driven Multistep Phase Transition of Charge Density Wave Order in 1T-TaS2},
  author={Yang, Qianyi and Shi, Wu and Zhong, Zhipeng and Li, Xiang and Li, Yan and Meng, Xiangjian and Wang, Jianlu and Chu, Junhao and Huang, Hai},
  journal={Nano Letters},
  volume={24},
  number={51},
  pages={16417--16425},
  year={2024},
  publisher={ACS Publications}
}

@article{mohammadzadeh2021evidence,
  title={Evidence for a thermally driven charge-density-wave transition in 1T-TaS2 thin-film devices: Prospects for GHz switching speed},
  author={Mohammadzadeh, Amirmahdi and Baraghani, Saba and Yin, Shenchu and Kargar, Fariborz and Bird, Jonathan P and Balandin, Alexander A},
  journal={Applied Physics Letters},
  volume={118},
  number={9},
  year={2021},
  publisher={AIP Publishing}
}

@article{ritschel2018stacking,
  title={Stacking-driven gap formation in layered 1 T-TaS 2},
  author={Ritschel, T and Berger, H and Geck, J},
  journal={Physical Review B},
  volume={98},
  number={19},
  pages={195134},
  year={2018},
  publisher={APS}
}

@inbook{Chaikin_Lubensky_1995, place={Cambridge}, title={Mean-field theory}, booktitle={Principles of Condensed Matter Physics}, publisher={Cambridge University Press}, author={Chaikin, P. M. and Lubensky, T. C.}, year={1995}, pages={144–212}}

@article{imada1998metal,
  title={Metal-insulator transitions},
  author={Imada, Masatoshi and Fujimori, Atsushi and Tokura, Yoshinori},
  journal={Reviews of modern physics},
  volume={70},
  number={4},
  pages={1039},
  year={1998},
  publisher={APS}
}

@book{kardar2007statistical,
  title={Statistical physics of fields},
  author={Kardar, Mehran},
  year={2007},
  publisher={Cambridge University Press}
}

@article{nayak1996possible,
  title={Possible electronic structure of domain walls in Mott insulators},
  author={Nayak, Chetan and Wilczek, Frank},
  journal={International Journal of Modern Physics B},
  volume={10},
  number={17},
  pages={2125--2136},
  year={1996},
  publisher={World Scientific}
}

@article{rawat2024symmetry,
  title={Symmetry breaking and structural instability in ultrathin 2 H-Ta S 2 across the charge density wave transition},
  author={Rawat, Divya and Thomas, Aksa and Singh Rana, Ajay Partap and Bera, Chandan and Soni, Ajay},
  journal={Physical Review B},
  volume={109},
  number={15},
  pages={155411},
  year={2024},
  publisher={APS}
}

@article{malarz2025confirming,
  title={Confirming universality of the fractal dimension of incipient percolation cluster for complex neighborhoods},
  author={Malarz, Krzysztof and Krawczyk, Malgorzata J},
  journal={Scientific Reports},
  volume={15},
  number={1},
  pages={32920},
  year={2025},
  publisher={Nature Publishing Group UK London}
}

@article{PhysRevLett.105.187401,
  title = {Ultrafast Melting of a Charge-Density Wave in the Mott Insulator $1T\mathrm{\text{\ensuremath{-}}}{\mathrm{TaS}}_{2}$},
  author = {Hellmann, S. and Beye, M. and Sohrt, C. and Rohwer, T. and Sorgenfrei, F. and Redlin, H. and Kall\"ane, M. and Marczynski-B\"uhlow, M. and Hennies, F. and Bauer, M. and F\"ohlisch, A. and Kipp, L. and Wurth, W. and Rossnagel, K.},
  journal = {Phys. Rev. Lett.},
  volume = {105},
  issue = {18},
  pages = {187401},
  numpages = {4},
  year = {2010},
  month = {Oct},
  publisher = {American Physical Society}
}

@article{singh2022lattice,
  title={Lattice-driven chiral charge density wave state in 1 T-TaS 2},
  author={Singh, Manoj and Yu, Boning and Huber, James and Sharma, Bishnu and Ainouche, Ghilles and Fu, Ling and Van Wezel, Jasper and Boyer, Michael C},
  journal={Physical Review B},
  volume={106},
  number={8},
  pages={L081407},
  year={2022},
  publisher={APS}
}

@article{salzmann2023observation,
  title={Observation of the metallic mosaic phase in 1 T-TaS 2 at equilibrium},
  author={Salzmann, Bj{\"o}rn and Hujala, Elina and Witteveen, Catherine and Hildebrand, Baptiste and Berger, Helmuth and von Rohr, Fabian O and Nicholson, Christopher W and Monney, Claude},
  journal={Physical Review Materials},
  volume={7},
  number={6},
  pages={064005},
  year={2023},
  publisher={APS}
}

@article{kuhn2019directional,
  title={Directional sub-femtosecond charge transfer dynamics and the dimensionality of 1T-TaS 2},
  author={K{\"u}hn, Danilo and M{\"u}ller, Moritz and Sorgenfrei, Florian and Giangrisostomi, Erika and Jay, Raphael M and Ovsyannikov, Ruslan and M{\aa}rtensson, Nils and S{\'a}nchez-Portal, Daniel and F{\"o}hlisch, Alexander},
  journal={Scientific reports},
  volume={9},
  number={1},
  pages={488},
  year={2019},
  publisher={Nature Publishing Group UK London}
}

@article{svetin2017three,
  title={Three-dimensional resistivity and switching between correlated electronic states in 1T-TaS2},
  author={Svetin, Damjan and Vaskivskyi, Igor and Brazovskii, Serguei and Mihailovic, Dragan},
  journal={Scientific reports},
  volume={7},
  number={1},
  pages={46048},
  year={2017},
  publisher={Nature Publishing Group UK London}
}

@article{park2023stacking,
  title={Stacking and spin order in a van der Waals Mott insulator 1T-TaS2},
  author={Park, Jae Whan and Lee, Jinwon and Yeom, Han Woong},
  journal={Communications Materials},
  volume={4},
  number={1},
  pages={99},
  year={2023},
  publisher={Nature Publishing Group UK London}
}

@article{yoshida2019charge,
  title={Charge density wave dynamics in nonvolatile current-induced phase transition in 1 T-Ta S 2},
  author={Yoshida, Masaro and Sato, Takuro and Kagawa, Fumitaka and Iwasa, Yoshihiro},
  journal={Physical Review B},
  volume={100},
  number={15},
  pages={155125},
  year={2019},
  publisher={APS}
}

@article{yin2024real,
  title={Real-Time Observation of Slowed Charge Density Wave Dynamics in Thinned 1T-TaS2},
  author={Yin, Shenchu and He, Keke and Barut, Bilal and Randle, Michael D and Dixit, Ripudaman and Nathawat, Jubin and Adinehloo, Davoud and Perebeinos, Vasili and Han, Jong E and Bird, Jonathan P},
  journal={Advanced Physics Research},
  volume={3},
  number={9},
  pages={2400033},
  year={2024},
  publisher={Wiley Online Library}
}

@article{roy2025metal,
  title={Metal to insulator crossover in the repulsive Fermi-Hubbard model probed by static correlations},
  author={Roy, Sayantan and Pervaiz, Sameed and Paiva, Thereza and Trivedi, Nandini},
  journal={Physical Review B},
  volume={112},
  number={16},
  pages={165144},
  year={2025},
  publisher={APS}
}

@article{wang2025fractionalization,
  title={Fractionalization signatures in the dynamics of quantum spin liquids},
  author={Wang, Kang and Feng, Shi and Zhu, Penghao and Chi, Runze and Liao, Hai-Jun and Trivedi, Nandini and Xiang, Tao},
  journal={Physical Review B},
  volume={111},
  number={10},
  pages={L100402},
  year={2025},
  publisher={APS}
}

@article{PhysRevB.54.R3756,
  title = {Superconductor-insulator transition in a disordered electronic system},
  author = {Trivedi, Nandini and Scalettar, Richard T. and Randeria, Mohit},
  journal = {Phys. Rev. B},
  volume = {54},
  issue = {6},
  pages = {R3756--R3759},
  numpages = {0},
  year = {1996},
  month = {Aug},
  publisher = {American Physical Society},
 }

@article{PhysRevLett.93.126401,
  title = {Inhomogeneous Metallic Phase in a Disordered Mott Insulator in Two Dimensions},
  author = {Heidarian, Dariush and Trivedi, Nandini},
  journal = {Phys. Rev. Lett.},
  volume = {93},
  issue = {12},
  pages = {126401},
  numpages = {4},
  year = {2004},
  month = {Sep},
  publisher = {American Physical Society},
}

@article{grover2010weak,
  title={Weak Mott insulators on the triangular lattice: possibility of a gapless nematic quantum spin liquid},
  author={Grover, Tarun and Trivedi, N and Senthil, T and Lee, Patrick A},
  journal={Physical Review B—Condensed Matter and Materials Physics},
  volume={81},
  number={24},
  pages={245121},
  year={2010},
  publisher={APS}
}

@article{morozovska2023landau,
  title={Landau-Ginzburg theory of charge density wave formation accompanying lattice and electronic long-range ordering},
  author={Morozovska, Anna N and Eliseev, Eugene A and Gopalan, Venkatraman and Chen, Long-Qing},
  journal={Physical Review B},
  volume={107},
  number={17},
  pages={174104},
  year={2023},
  publisher={APS}
}

@article{mcmillan1975landau,
  title={Landau theory of charge-density waves in transition-metal dichalcogenides},
  author={McMillan, WL},
  journal={Physical Review B},
  volume={12},
  number={4},
  pages={1187},
  year={1975},
  publisher={APS}
}

@article{moura2024mcmillan,
  title={McMillan-Ginzburg-Landau theory of singularities and discommensurations in charge density wave states of transition metal dichalcogenides},
  author={Moura, VN and Chaves, A and Peeters, FM and Milo{\v{s}}evi{\'c}, MV},
  journal={Physical Review B},
  volume={109},
  number={9},
  pages={094507},
  year={2024},
  publisher={APS}
}

@article{nakatsugawa2020multivalley,
  title={Multivalley free energy landscape and the origin of stripe and quasi-stripe CDW structures in monolayer MX2 compounds},
  author={Nakatsugawa, Keiji and Tanda, Satoshi and Ikeda, Tatsuhiko N},
  journal={Scientific Reports},
  volume={10},
  number={1},
  pages={1239},
  year={2020},
  publisher={Nature Publishing Group UK London}
}

@article{domrose2023light,
  title={Light-induced hexatic state in a layered quantum material},
  author={Domr{\"o}se, Till and Danz, Thomas and Schaible, Sophie F and Rossnagel, Kai and Yalunin, Sergey V and Ropers, Claus},
  journal={Nature materials},
  volume={22},
  number={11},
  pages={1345--1351},
  year={2023},
  publisher={Nature Publishing Group UK London}
}

@article{bar2018kinetic,
  title={Kinetic spinodal instabilities in the mott transition in V2O3: Evidence from hysteresis scaling and dissipative phase ordering},
  author={Bar, Tapas and Choudhary, Sujeet Kumar and Ashraf, Md Arsalan and Sujith, KS and Puri, Sanjay and Raj, Satyabrata and Bansal, Bhavtosh},
  journal={arXiv preprint arXiv:1808.00693},
  year={2018}
}

@article{klein1954principle,
  title={Principle of minimum entropy production},
  author={Klein, Martin J and Meijer, Paul HE},
  journal={Physical Review},
  volume={96},
  number={2},
  pages={250},
  year={1954},
  publisher={APS}
}

@book{Prigogine1947,
  author       = {Prigogine, Ilya},
  title        = {Étude thermodynamique des phénomènes irréversibles},
  year         = {1947},
  publisher    = {Desoer / Dunod},
  address      = {Liège / Paris},
  note         = {Original work formulating the minimum entropy production theorem},
}

@article{guo2017thermodynamic,
  title={Thermodynamic extremum principles for nonequilibrium stationary state in heat conduction},
  author={Guo, Yangyu and Wang, Ziyan and Wang, Moran},
  journal={Journal of Heat Transfer},
  volume={139},
  number={7},
  pages={071303},
  year={2017},
  publisher={American Society of Mechanical Engineers}
}

@article{BertolaCafaro2008,
  author    = {V. Bertola and E. Cafaro},
  title     = {A critical analysis of the minimum entropy production theorem and its application to heat and fluid flow},
  journal   = {International Journal of Heat and Mass Transfer},
  volume    = {51},
  number    = {7-8},
  pages     = {1907--1912},
  year      = {2008},
  doi       = {10.1016/j.ijheatmasstransfer.2007.06.041},
}

@article{chen2009thermal,
  title={Thermal contact resistance between graphene and silicon dioxide},
  author={Chen, Z and Jang, W and Bao, W and Lau, CN and Dames, C},
  journal={Applied Physics Letters},
  volume={95},
  number={16},
  year={2009},
  publisher={AIP Publishing}
}

@misc{Supple,
 title = {See supplementary details at   for X-ray photo emission spectroscopy, thickness dependent study, device frabrications and measurements details, which includes Refs. [12, 37, 45, 64, 73, 79-81, 83, 86, 91, 92]}
}

@article{stojchevska2014ultrafast,
  title={Ultrafast switching to a stable hidden quantum state in an electronic crystal},
  author={Stojchevska, L and Vaskivskyi, I and Mertelj, T and Kusar, P and Svetin, D and Brazovskii, S and Mihailovic, D},
  journal={Science},
  volume={344},
  number={6180},
  pages={177--180},
  year={2014},
  publisher={American Association for the Advancement of Science}
}

@article{cho2016nanoscale,
  title={Nanoscale manipulation of the Mott insulating state coupled to charge order in 1 T-TaS2},
  author={Cho, Doohee and Cheon, Sangmo and Kim, Ki-Seok and Lee, Sung-Hoon and Cho, Yong-Heum and Cheong, Sang-Wook and Yeom, Han Woong},
  journal={Nature communications},
  volume={7},
  number={1},
  pages={10453},
  year={2016},
  publisher={Nature Publishing Group UK London}
}

\pagebreak
\begin{figure}[h!]
\centerline{\includegraphics[scale=0.75, clip] {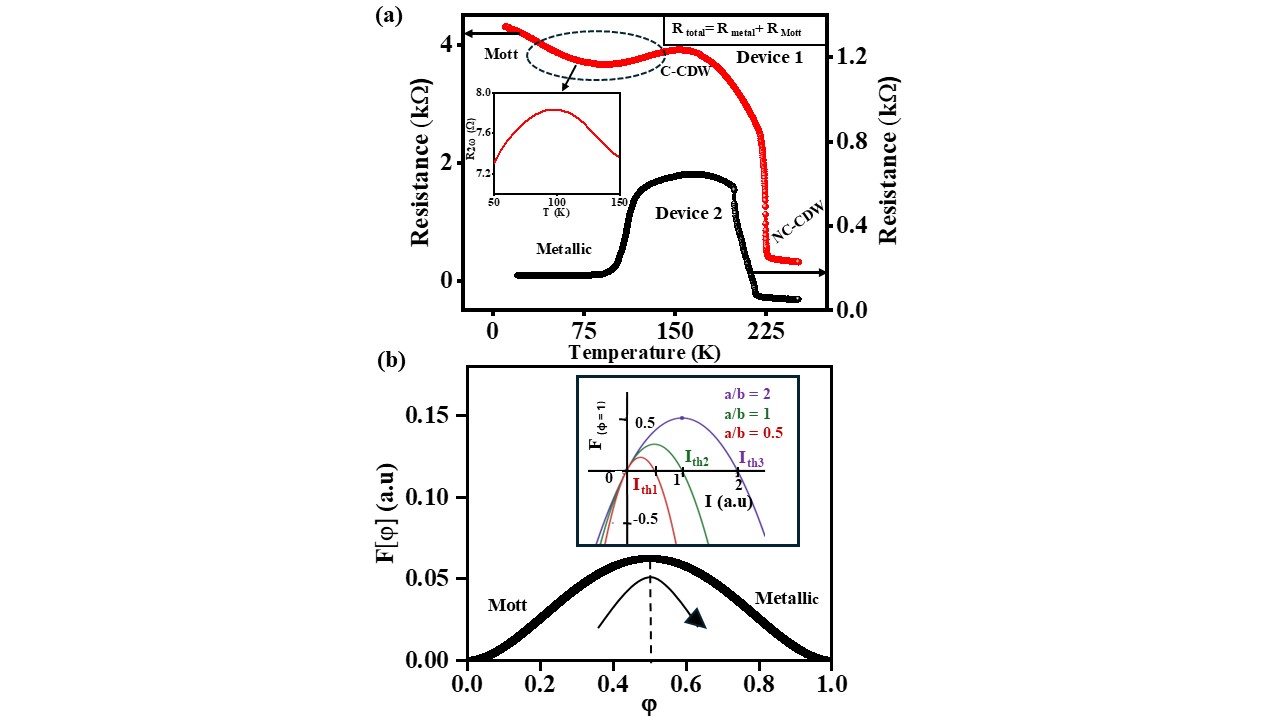}}
\caption{\textbf{Temperature-dependent resistance ($R$) and free-energy landscape $F(\phi)$.} 
\textbf{(a)} $R$–$T$ for two flakes of differing thickness, showing low-$T$ Mott (red) and metallic (black) ground states. 
\textit{Inset:} 2$\omega$ response of resistance (50–150 K) for the Mott device, where a resistance dip coincides with a Gaussian-like 2$\omega$ peak, indicating structural reorientation and coexistence of metallic and Mott fractions. 
\textbf{(b)} At $I=0$, $F(\phi)$ exhibits a double well with minima at $\phi=0$ (Mott) and $\phi=1$ (metallic), separated by a maximum at $\phi=0.5$. 
\textit{Inset:} $I$  shifts the endpoint energy $F(\phi=1)$ set by $a/b$ ($a$: scattering coefficient, $b$: $R\times t$). 
Above $I_{\mathrm{th}}$, the metallic state is energetically favored.}


\end{figure}

\pagebreak
\begin{figure}[h!]
\centerline{\includegraphics[scale=0.9, clip]{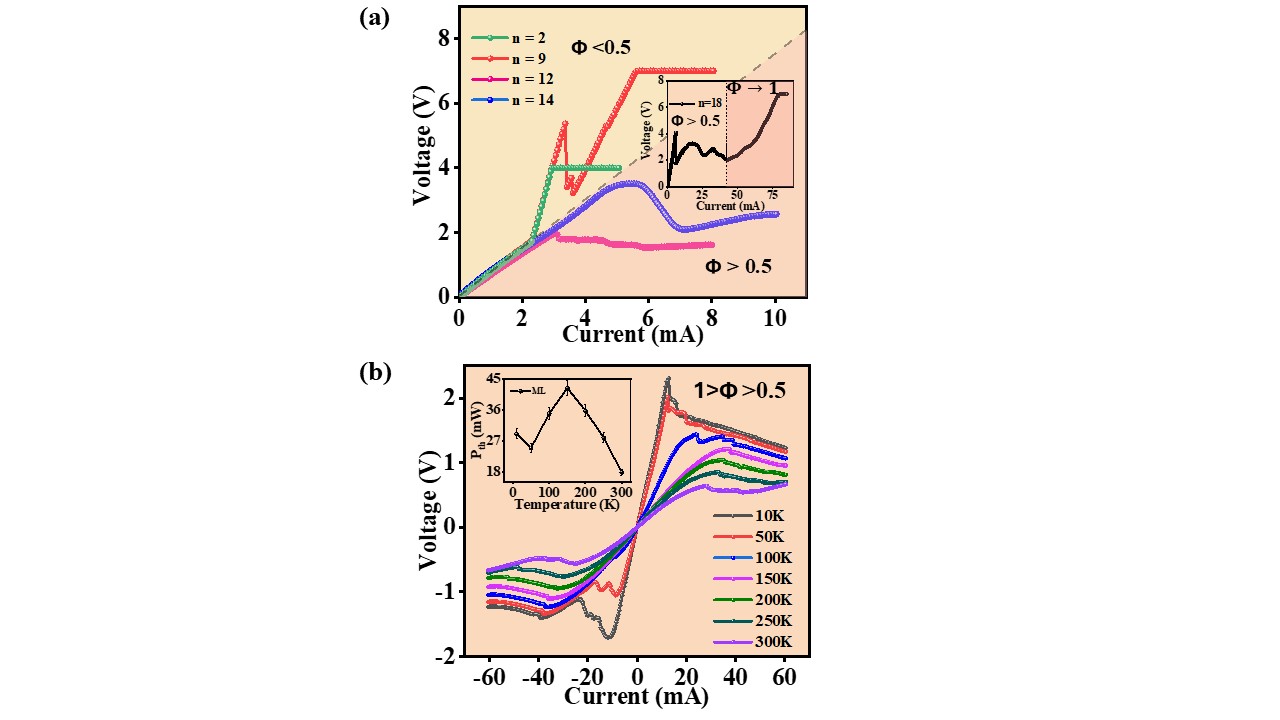}}
\caption {\textbf{Voltage–current characteristics of MI and ML states.} \textbf{(a)} $V$–$I$ of the Mott-insulating (MI) state at 10 K for successive current loops ($n$), showing a gradual transition from $\phi=0$ to $\phi=1$ with the emergence of pronounced negative differential resistance (NDR). \textbf{(b)} $V$–$I$ of the metal-like (ML) state from 10–300 K. At low $T$, strong NDR accompanies current-induced transitions to highly conductive states; with increasing $T$, NDR sharpness diminishes, indicating reduced resistance-switching dynamics. Inset: threshold power for NDR vs $T$, reflecting the temperature-dependent coupling between current-driven percolation and structural instabilities.}

\end{figure}

\pagebreak
\begin{figure}[h!]
\centerline{\includegraphics[scale=0.65, clip]{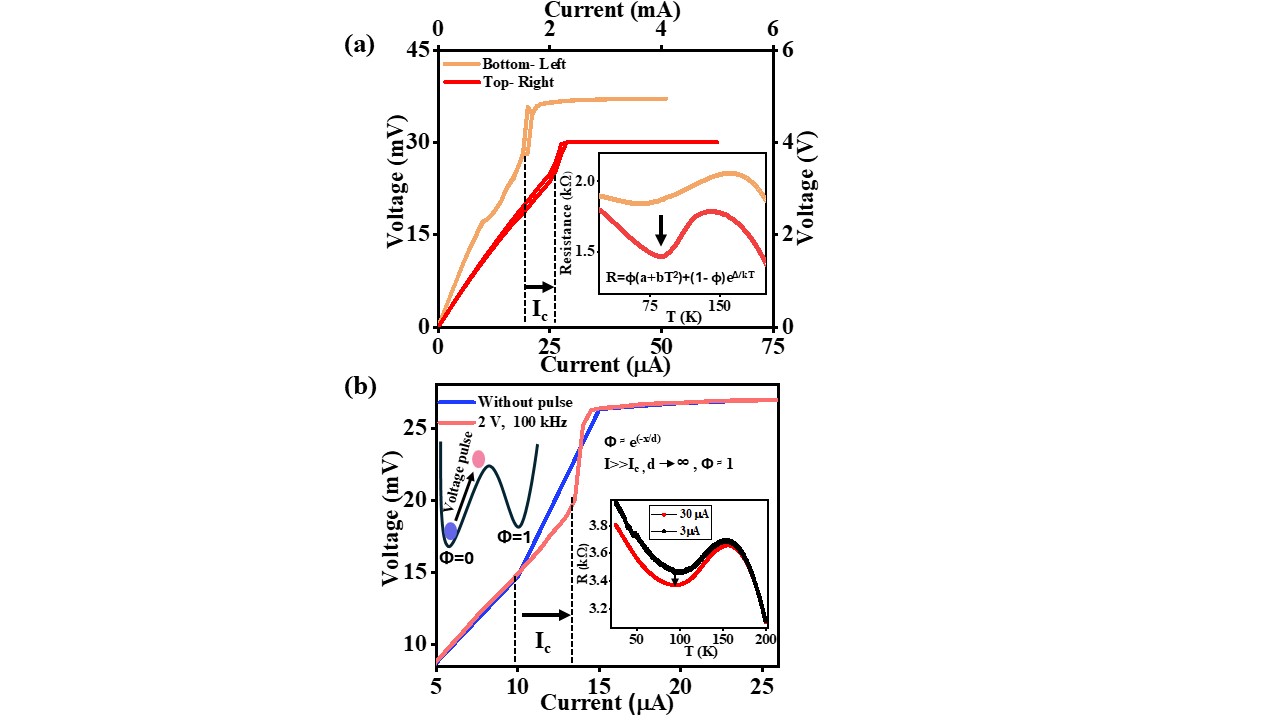}}
\caption{\textbf{Device-dependent responses and tuning of the $\phi=0 \rightarrow 1$ transition.} 
\textbf{(a)} $V$–$I$ and $R$–$T$ for two devices (yellow, red) with differing metallic fractions. 
The critical current $I_c$ (where $R$ jumps) correlates with the $R$–$T$ dip (50–150~K), with larger dips yielding higher $I_c$, indicating a link between $I_c$ and order parameter $\phi$ for $I\leq0.5$. No NDR is observed in this regime. 
\textbf{(b)} Effect of voltage pulses: without pulse (blue) and after a 2~V, 100~kHz, 60~s pulse (pink). 
Pulse application shifts $I_c$ higher for $I\leq0.5$, consistent with increased metallic fraction. 
\textit{Inset:} $R$–$T$ for the same device at $I=3~\mu$A and $30~\mu$A; larger currents deepen the $R$–$T$ dip, indicating current-induced growth of the metallic fraction.}

\end{figure}

\pagebreak
\begin{figure}[h!]
\centerline{\includegraphics[scale=0.65, clip]{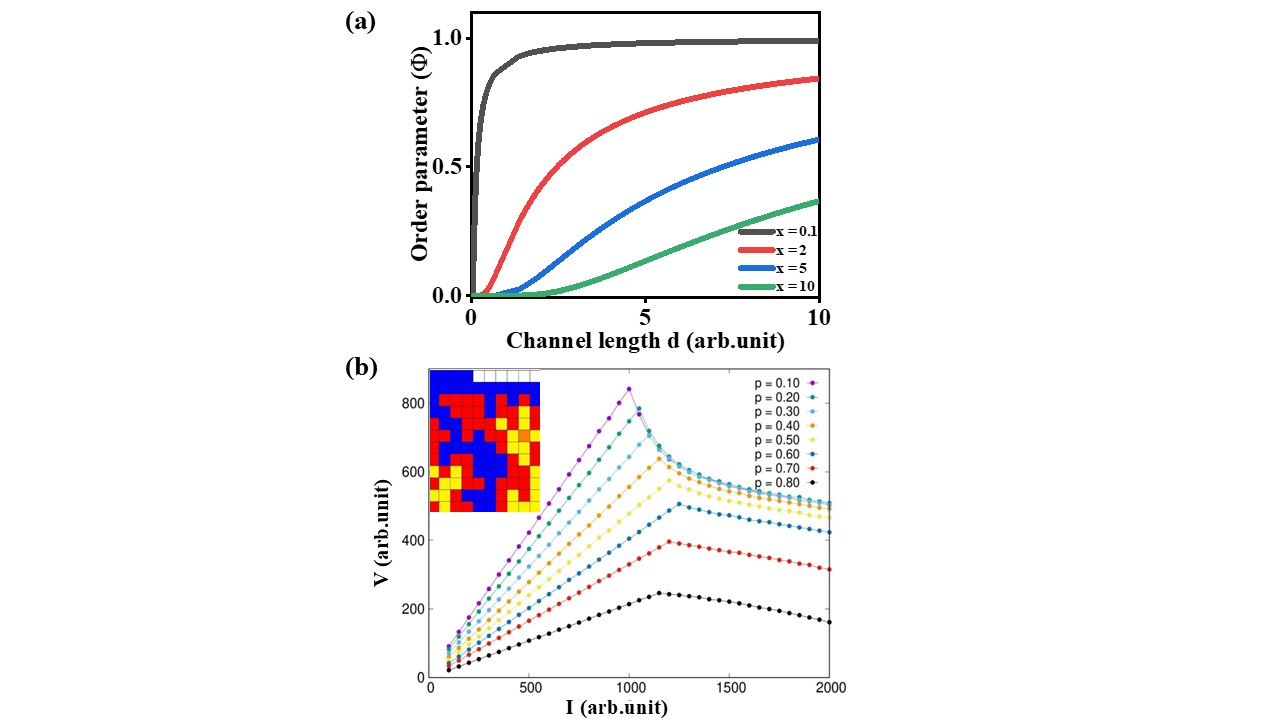}}
\caption{\textbf{(a)} Characteristics of order parameter ($\phi$) with channel length (d) ($\phi$ \textit{vs.} d), for different values of domain channel ratio. Here x/d is the domain to channel ratio, for smaller value of x, $\phi$ easily approach 1 with variation of d, as x increases sufficiently $\phi$ can never reach 1 (metallic state). \textbf{(b)} Calculated \( V-I \) characteristics of the Metallic state for varying \( p_c \). The Metallic state demonstrates a lower resistance and smoother transition, indicating the formation of continuous conductive channels. The inset illustrates the percolation model: red denotes closed (blocked) sites, yellow represents open (waiting) sites, and blue highlights the continuous channels formed by the connected open sites. This schematic emphasizes the role of percolation dynamics in the emergence of conductive pathways across the system, as described in the text.  
}
\label{VI_perco}
\end{figure}

\end{document}


\title{ Supporting Information for:
\newline Percolative Instabilities and Sparse-Limit Fractality in 1T-TaS$_2$}
\author{Poulomi Maji}
\affiliation{School of Physical Sciences, Indian Association for the Cultivation of Science, 2A $\&$ B
Raja S. C. Mullick Road, Jadavpur, Kolkata - 700032, India
}

\author{Md Aquib Molla}
\affiliation{Vidyasagar College, Department of Physics, 39 Sankar Ghosh Lane, Kolkata 700006,India
}

\author{Koushik Dey}
\affiliation{School of Physical Sciences, Indian Association for the Cultivation of Science, 2A $\&$ B
Raja S. C. Mullick Road, Jadavpur, Kolkata - 700032, India
}

\author{Bikash Das}
\affiliation{School of Physical Sciences, Indian Association for the Cultivation of Science, 2A $\&$ B
Raja S. C. Mullick Road, Jadavpur, Kolkata - 700032, India
}
\author{Sambit Choudhury}
\affiliation{UGC-DAE Consortium for Scientific Research, Khandwa Road, Indore 452001, Madhya Pradesh, India}

\author{Tanima Kundu}
\affiliation{School of Physical Sciences, Indian Association for the Cultivation of Science, 2A $\&$ B
Raja S. C. Mullick Road, Jadavpur, Kolkata - 700032, India
}

\author{Pabitra Kumar Hazra}
\affiliation{School of Physical Sciences, Indian Association for the Cultivation of Science, 2A $\&$ B
Raja S. C. Mullick Road, Jadavpur, Kolkata - 700032, India
}

\author{Mainak Palit}
\affiliation{School of Physical Sciences, Indian Association for the Cultivation of Science, 2A $\&$ B
Raja S. C. Mullick Road, Jadavpur, Kolkata - 700032, India
}

\author{Sujan Maity}
\affiliation{School of Physical Sciences, Indian Association for the Cultivation of Science, 2A $\&$ B
Raja S. C. Mullick Road, Jadavpur, Kolkata - 700032, India
}
\author{Bipul Karmakar}
\affiliation{School of Physical Sciences, Indian Association for the Cultivation of Science, 2A $\&$ B
Raja S. C. Mullick Road, Jadavpur, Kolkata - 700032, India
}
\author{Kai Rossnagel}
\affiliation{Ruprecht Haensel Laboratory, Deutsches Elektronen-Synchrotron DESY, Notkestr. 85, 22607 Hamburg, Germany}
\affiliation{Institut für Experimentelle und Angewandte Physik, Christian-Albrechts-Universität zu Kiel, Olshausenstr. 40, 24098 Kiel, Germany}

\author{Sanjoy Kr Mahatha}
\affiliation{UGC-DAE Consortium for Scientific Research, Khandwa Road, Indore 452001, Madhya Pradesh, India}
\author{Bhaskaran Muralidharan}
\affiliation{Indian Institute of Technology Bombay, Mumbai - 400076, India}

\author{Shamashis Sengupta}
\affiliation{Université Paris-Saclay, CNRS/IN2P3, IJCLab, 91405 Orsay, France}

\author{Sanchari Goswami}
\affiliation{Vidyasagar College, Department of Physics, 39 Sankar Ghosh Lane, Kolkata 700006,India
}

\author{Subhadeep Datta}
\email{sspsdd@iacs.res.in}
\affiliation{School of Physical Sciences, Indian Association for the Cultivation of Science, 2A $\&$ B
Raja S. C. Mullick Road, Jadavpur, Kolkata - 700032, India
}
\maketitle

\section{1T-TaS$_{2}$ Star of David structure (SOD)}
\textbf{X-ray photo emission spectroscopy}
Temperature-dependent Ta 4\emph{f} core-level spectra on \emph{in-situ} cleaved  1T-TaS$_{2}$ single-crystal were measured at the ASPHERE III end-station at beamline P04 of PETRA III, DESY using a photon energy of 260 eV. The total energy resolution was better than 80 meV. The evolution of the CDW split Ta 4\emph{f} core-level spectra while warming up the sample is shown in Fig.~S1(a).   
Lowering the sample temperature leads to the formation of a Star of David (SOD) structure, which induces a distinct electron distribution of tantalum (Ta) atoms in the system. In this arrangement, 12 Ta atoms contract towards the central atom, creating an electron density gradient between the A, B, and C types of Ta atoms, as shown in Fig.~S1(b). The C-type Ta atoms transfer 0.4 electrons to both A and B types, while B-type Ta atoms also transfer electrons towards the central atom. This results in three distinct types of Ta atoms based on their electron density variations. While the electron density differences between A and B types of Ta atoms are not detectable by X-ray photoelectron spectroscopy (XPS), differences between the C and B types cause a splitting of the $4f_{7/2}$ and $4f_{5/2}$ peaks below a certain temperature threshold, as the SOD structure forms \cite{PhysRevLett.105.187401}.

Below 550K the Ta atom of 1T-TaS2 starts to distort and the lattice parameters modifies to minimize the overall energy of the system and forms incommensurate charge density wave (IC-CDW), below 350K these 12 Ta atoms comes closer to the central Ta atoms and forms hexagonal domains Star of David structure (SOD) (see Fig.~S1(b) (iv)) forming nearly commensurate state (NC-CDW) separated by incommensurate domain walls. Below 180K, fully locked periodic modulation of electron density occurs and CDW wave vector becomes commensurate with the crystal lattice. $\surd$13x$\surd$13 superstructure of SOD forms leading to commensurate CDW (C-CDW), where the central Ta atom of each SOD has one unpaired electron \cite{PhysRevB.100.155407,ritschel2015orbital}. The out-of-plane arrangement of these SOD below the C-CDW temperature can give rise to a Mott insulating state or a metallic state, depending on the stacking configuration.  Two distinct types of out-of-plane stacking of this SOD, namely AL and L as presented in Fig.~S1(b)(i, ii), govern the low-temperature state. In AL stacking two adjacent layers forms t$_0$ dimer(Fig.~S1(b) (iii)), where the Star of David (SOD) structure is exactly on top of each other hence the unpaired electron of the central Ta atoms are strongly correlated, with no lateral shift (A-stacking) between them, followed by a lateral shift to the next pair presented in Fig.~S1(b). Due to this bilayer dimerization, a gap opens up, making the system a Mott insulator. In the L stacking, David stars in adjacent layers are laterally shifted or rotated, often randomly, hence SODs are not exactly on top of each other, central Ta atom’s electron are not anymore strongly correlated, free to conduct, gives rise to a metallic response in the system . Most samples exhibit predominant AL stacking with intermittent L stacking, resulting in a mixed Mott-insulating and metallic phase.

\section{Electron localization function of the room temperature and low temperature supercell structure}
To visualize the electron distribution of Ta atoms of the room temperature NC-CDW and low temperature C-CDW state, we calculated the electron localization function (ELF). Room temperature NC-CDW state's Ta atoms has no electronic overlap in the out of plane direction or in plane direction presented in Fig. ~S2(a), Fig.~S2(c) respectively. This explains the metallic like electron conduction in the channel dominated NC-CDW state. While in the low temperature below 220 K, $\sqrt{13}\times\sqrt{13}$ superstructure is formed. Beyond 100 K due to strong out of plane electronic correlation (AL stacking) between the central Ta atoms of the SOD, resistivity takes upturn with temperature showing the Mott nature. This has been captured through the ELF, in the out of plane direction there is substantial electronic overlap between the central A type Ta atoms of the SOD across the Van der Waals gap presented in Fig.~S2(b). In the in-plane direction B and C types Ta atoms of the supercell have huge electronic overlap (Fig.~S2(d)), explains the electronic correlation of the Mott state. Previous finding suggests reducing thickness, rapid cooling, femto second laser pulse, voltage pulse suppress the NC-CDW to C-CDW transition and stabilizes a metastable NC-CDW state \cite{stojchevska2014ultrafast, yoshida2014controlling, yoshida2015memristive,vaskivskyi2015controlling,stahl2020collapse}. Low temperature hidden metallic state thus has been termed as the supercooled NC-CDW state due to close structural similarities with the room temperature NC-CDW state, hence follows simillar electronic distribution ( not much electronic overlap between the out of plane, in-plane  Ta atoms). Our electron localization function calculations thus indirectly provides an idea about the difference in electronic distribution between the low temperature Mott and metallic state of the system.   

\section{Thickness dependence of metallic state}
Reducing thickness of flake below a critical thickness of approximately 7–8 layers,leads to the reduction in interlayer coupling (disruption of the correlation between the electrons of central Ta atoms of t$_0$ bilayer SOD) which suppresses the Mott insulating state. However, emergence of metallicity is not solely dictated by thickness. Rather, the key factor is the disruption of the out-of-plane CDW stacking order. Stacking misalignments (L stacking> AL stacking) can destabilize the commensurate CDW phase, leading to a collapse of the Mott gap. Meaning naturally if L stacking is more dominant than AL stacking, the system will show metallic behavior below the Commensurate charge density wave (C-CDW) temperature. Consequently, even in thicker samples (e.g., thickness $>$20 nm), metallic behavior below 180K can be favored, if stacking disorder is present. As a result for both devices with thickness 60nm and 20nm, showed low-temperature metallic behavior as presented in Fig.~S3(a). Also in some cases even in thinner sample (15-25) nm, Mott insulating state prevealed at low temperature. So lower thickness can promote metallicity to emerge, along with different perturbations such as current, voltage pulse, ionic liquid gating etc, but does not guarantee metallicity as stacking alteration is the main driving force.

\begin{table}[h!]

\caption{Statistical distribution of 30 flakes showing two distinct low-temperature behaviors — Mott insulating and metallic — with thickness variation. }
\begin{tabular}{ccc} 
\toprule
Number of devices & Thickness of flakes & Low temperature nature of sample \\ 
\midrule
17 & (40-60) nm & Mott \\
5 & (10-25) nm & Mott \\
6 & (5-25) nm & Metallic \\
2 & (40-60) nm & Metallic \\
\bottomrule
\end{tabular}
\end{table}

\section{Second harmonic response of resistance  with temperature of a low temperature Mott device}
Second harmonic (2$\omega$) measurements were performed on low-temperature Mott device of Fig. 1(a) using lock-in amplifier SR830 DSP, with input ac current of 50$\mu$A, 27.777Hz frequency. A voltage amplifier has been used to detect proper second harmonic signal. The 2$\omega$ response of the resistance exhibits a sharp peak at 225~K, corresponding to the transition from the commensurate charge density wave (C-CDW) to the nearly commensurate CDW (NC-CDW) phase, indicating a structural reorganization of the system, presented in Fig.~S4(a) for heating cycle. Interestingly, a broad Gaussian-like peak appears in the temperature range of 50~K to 150~K, which correlates with the dip observed in the $R$–$T$ curve within the same range. This suggests a structural reconfiguration driven by the competition between two types of stacking configurations: 'AL' (associated with the Mott state) and 'L' (associated with the metallic state), hence at low temperature system is combination of both types of stacking. The overall resistance ($R_{total}$) can thus be modeled as a mixture of metallic ($R_{ML}$$\sim {T^2}$) and Mott-like behavior ($R_{MI}$ $\sim e^{\Delta / k_{B}T}$):
\begin{equation}
    R_{\text{total}} = \phi(\alpha + \beta T^2) + (1 - \phi) e^{\Delta / k_{B}T},
\end{equation} Presented in Fig.~S4(b). The system settles into either the Mott insulating or metallic ground state at low temperatures, depending on the dominant stacking configuration.

\section{Voltage-Current response of Mott device deep in $\phi$ = 0 valley}

We investigated the current-driven nonlinear dynamics of another Mott device with higher low temperature resistance around 2 k$\Omega$ (V-I). At low current bias, the V-I curve is linear, as the applied current increases, the system transitions to a low-resistance branch. For intermediate currents, the V-I curve evolves and connects these two branches. The key question is to determine which path the system follows as it transitions between these branches. Three distinct possibilities emerge, presented in Fig.~S6(b). The first is a path with \( \frac{dV}{dI} > 0 \) (yellow curve), which suggests that the system is fragmented into domains of high- and low-resistance phases, with the higher-resistance domains exhibiting insulating behaviour while the lower-resistance domains facilitate current conduction. In this case, the positive slope indicates that the system's resistance increases as the current flows through a more resistive domain network. The second possibility is a path with \( \frac{dV}{dI} < 0 \) (pink curve), implying that the system fragments into domains with a higher proportion of the better-conducting, high-temperature phase. Here, the negative slope suggests that the system undergoes a transition to a state where conduction is dominated by the more conductive domains, reducing the overall resistance as current is applied. Alternatively, an intermediate branch where \( \frac{dV}{dI} = 0 \) (green curve) could also be considered. This represents a critical state where the system is in a balance, with no net voltage change in response to current variations. Physically, this could indicate a steady state in which the conduction channels exhibit a uniform distribution of resistive phases, resulting in a stable, non-volatile state \cite{PhysRevLett.87.126401, kumar2017physical, mingtao2006negative}.

Starting from the MI state 10 K, shown in Fig.~S6(a), a 20 $\mu$A current was sufficient to induce a transition from a higher-resistance state (1.8 k$\Omega$) to a lower-resistance state (1.2 k$\Omega$), with resistance continuing to decrease as current increased. At the transition point, \( \frac{dV}{dI} \) approached zero, and the branch with \( \frac{dV}{dI} > 0 \) was absent, suggesting that the system underwent a phase transition where the system’s internal structure reorganized without significant change in the overall resistance. The voltage remained nearly constant, as the increase in current was balanced by a reduction in resistance in domains. Upto 190 K $I_{\text{th}}$ and resistance remain constant presented in Fig.~S6(a). The minimal resistance change suggests, along with sudden increase in resistance suggests CDW scattering across the domain boundary forming a thermal bottleneck equivalent to I = 0.5 ( Fig.~S4(a)) region means further current required to transition to the metallic state ($\phi$ = 1). This points to a subtle rearrangement of the system’s electronic structure rather than a dramatic phase change, no NDR was observed, and the \( \frac{dV}{dI} < 0 \) branch was absent, likely because domain fragmentation had not yet occurred. This absence of NDR (see Fig.~S8(d))further suggests that the system had not reached the conditions required for a percolative breakdown of the CDW phase, which is associated with where the system lies in the free energy landscape \cite{PhysRevLett.87.126401, kumar2017physical, mingtao2006negative}. Above 220 K, initial resistance drops from 1.8 k$\Omega$ to 170 $\Omega$ before transitioning to 120 $\Omega$ after a critical current of 188 $\mu$A (Fig.~S6(a)). Above 190 K, threshold power increases linearly, with very minimal NDR observed up to 250 K. The constant $I_{\text{th}}$ up to 190 K suggests stable domains, while by 220 K, the initial resistance decreased, and current driven transition towards lower resistance through the sudden surge in resistance, suggests system moving towards $\phi$ $>$ 0, but current still in I = 0.5 region, means further current required to transition to the metallic state ($\phi$ = 1). This points to a subtle rearrangement of the system’s electronic structure rather than a dramatic phase change, no NDR was observed, and the \( \frac{dV}{dI} < 0 \) branch was absent, likely because domain fragmentation had not yet occurred. This absence of NDR further suggests that the system had not reached the conditions required for a percolative breakdown of the CDW phase, which is associated with where the system lies in the free energy landscape \cite{PhysRevLett.87.126401, kumar2017physical, mingtao2006negative}.

\section{Resistor network model: explains low temperature metallic device voltage-current characteristics of Fig.~2(b).}
To explain the Voltage-Current response of the low temperature metallic device corresponding to Fig.~2(b) a schematic representation is presented in Fig.~S9 which illustrates the variable resistor network. Under an applied current with sweep rate $\tau_S$, an intermediate state nucleates, characterized by lower resistance through partial spatial nucleation of a percolated ML state. As the current increases, the nucleated region expands over a characteristic timescale $\tau_{\text{NUC}} > \tau_S$. Consequently, during the forward sweep, resistance changes respond on this longer timescale, causing resistance to drop only at currents exceeding the threshold nucleation current $I_{\text{th}}$. This delay results in a voltage increase until $\tau_{\text{NUC}}$ is reached, followed by a sharp resistance drop, producing sweep rate-dependent negative differential resistance (NDR) observed in the central figure (see Fig.~S9). In the backward sweep, the high-resistance state returns at $I_{\text{th}}$, creating the hysteresis (see Fig.~S8(a)). The differential conductance (dV/dI) measurements of the low temperature metallic state voltage-current response (Fig. 2(b)) at 10 K and 300 K indicated in blue and black color respectively, is  presented in Fig.~S8(b). The circuit diagram corresponding to dV/dI is presented in Fig.~S8(c).  

\section{Calculation of Joule heating}
For both MI and ML states, threshold power (P$_{th}$) remains in the mW range, insufficient to reach the C-CDW to NC-CDW transition temperature.Following the famous heat balanced equation P=$\gamma$$\Omega(T^6_e-T^6_{ph})$, $\gamma$= electron -phonon coupling strength in the unit of W/$K^6 m^3$. For disordered insulators, $\gamma = 1.85$~nW/$\mu$m$^3$K$^6$; 1T-TaS$_2$ is known to exhibit stronger coupling, giving $\gamma \sim 10^7$–$10^8$~W/K$^6$·m$^3$.. For 10K I-V at the transition point for both metallic and Mott, joule heating was 24mW-40mW respectively. Now considering at transition point electron and phonon decouples due to Joule heating produced by current, the equation reduces to P=$\gamma$ $\Omega$ T$^6_e$, T$_{el}$ $>$ T$_{ph}$ for $\Omega$ = 40$\mu$m $\times$ 10nm$\times$ 40$\mu$m. Then T$_{el}$ approximately reduces to 30 K to 40 K which is far below the C-CDW to NC-CDW transition temperature, however elevated local temperature can cause thermal runaway, leading to avalanche breakdown of domains.
Also considering Kapitza resistance (thermal boundary resistance) due to the TaS\textsubscript{2}/SiO$_{2}$ interface, 
$P = G_{\text{eff}} (T_{\text{flake}} - T_{\text{bath}})$, where 
$G_{\text{eff}} = \left( \dfrac{K_{\text{sub}}}{t_{\text{sub}}} + \dfrac{1}{R_k} \right) \times \text{area of contact}$. Where $G_{\text{eff}}$ is the total effective thermal conductance of the interface, R$_k$ is the kapitza resistance ( resistance of the interface), K$_{sub}$ is the thermal conductance of SiO$_2$, t$_{sub}$ is the thickness of the substrate, 
$K_{\text{sub}} = 0.05~\text{W}/\text{m} \cdot \text{K}$, 
$t_{\text{sub}} = 300~\text{nm}$, 
and at low temperature $R_k \approx 10^{-7}~\text{m}^2 \cdot \text{K}/\text{W}$. 
Gives $T_{\text{flake}} = 20~\text{K}$ and bath temperature $T_{\text{bath}} = 10~\text{K}$, hence minimal overall increase of temperature of device can occur due to Joule heating in the mW range\cite{chen2009thermal}.

\section{Sample preparation and device fabrication}
\textbf{Sample preparation:}
As reported in the previous literature \cite{PhysRevLett.120.166401, wang2024dualistic,tsen2015structure}.

\textbf {Device fabrication and measurements:}
1T-TaS$_2$ flakes were exfoliated on clean Si substrate coated with 285 nm SiO$_2$. Suitable flakes were located using an optical microscope by comparing their contrast. Four probe devices were fabricated by electron beam / optical lithography, followed by Cr(10nm)-Au(80nm) evaporation. The difference between two voltage probe was 4 micron. Low temperature measurements were done in a closed cycle cryostat (ARS-4HW). All four probe  measurements were performed using 6220 precision current source meters and 2182 nanovoltmeters.
\newline
 The differential conductance measurements ($dI/dV$) as well as 2$\omega$ response of resistance was measured using lock-in amplifier SR830 DSP. Differential conductance measurements for the low resistance sample circuit diagram is given in Fig. S8(c). Ac current of 7 micro-ampere of frequency 733.37 Hz was superimposed with dc current. Dc output voltage part was measured via Keithley 2601B source meter and ac voltage was measured in the lock-in amplifier. Lock-in was set on A-B mode, to measure the difference in ac voltage of the 2 voltage terminals.
Several devices have been measured, only 10\% showed low-temperature metallic response naturally. Sudden increase in resistance has been consistent in Voltage-Current (V vs I) response across the Mott devices. Where, the critical current (I$_c$) for sudden increase in resistance and threshold current (I$_{th}$) required for NDR are different for different devices, depends on the position of system on the free energy landscape.  

 \section{Percolation}
 For percolation model, we considered a 2D block of $n \times n$ sites. Here two types of sites are considered : open sites and closed/blocked sites. Sites are open with probability $p$ and blocked with probability $(1-p)$. There exists a threshold value of $p$, i.e., $p_c$, for usual percolation which is $0.59$. Below $p_c$, the system is unable to percolate and above a $p_c$, percolation is allowed. At a particular $p$ below $p_c$, e.g., at $p=0.4$ most of the sites are blocked and only $40$\% sites are open. As huge number of the sites are blocked below $p_c$, the resistance is high (see Fig.~S10(a) inset). No percolating path can be found in that case. Therefore, if a current source is connected to the system, as the resistance is very high, the drop across the resistor (the system) will be high enough to measure.With increase in current, the drop across the model system will also increase almost proportionally. Thus the V-I graph will be a straight line with a positive slope presented in Fig. ~S10(a). However, beyond a certain value of current, due to local joule heating effect, more sites may be open within the system. This is something like the thundering effect in the Forest Fire Model (FFM) in percolation. For this, the resistance of the system will be lowered and the slope of the graph is diminished. This clearly mimics the low temperature state/ MOTT state of the system.For $p$ above $p_c$, there are a large number of open sites. Therefore the resistance of the system is low and it can percolate. A large amount of current is required to measure an appreciable amount of voltage drop across the system as presented in Fig.~S10(b). As for high $p$ due to increased temperature (T) we are probing the system with high current, there will be chances of more open sites due to heating. Therefore, the effective $p$ is higher than the initial $p$ value we are starting with. The slopes of the curves get lowered with increase in $T$ for the original system and increase in $p$ in the percolation model system. This high $p$ picture is thus equivalent to the high $T$ state/NC-CDW state of layered $1T-TaS_{2}$ presented in Fig.~S10(b). The threshold $p$. i.e. $p_c$ here may be considered as the transition temperature $T \sim 220K$ for the Mott insulating state.
 \\

\bibliography{refsupple}

\newpage

\begin{figure}[h!]
\centerline{\includegraphics[scale=0.65, clip]{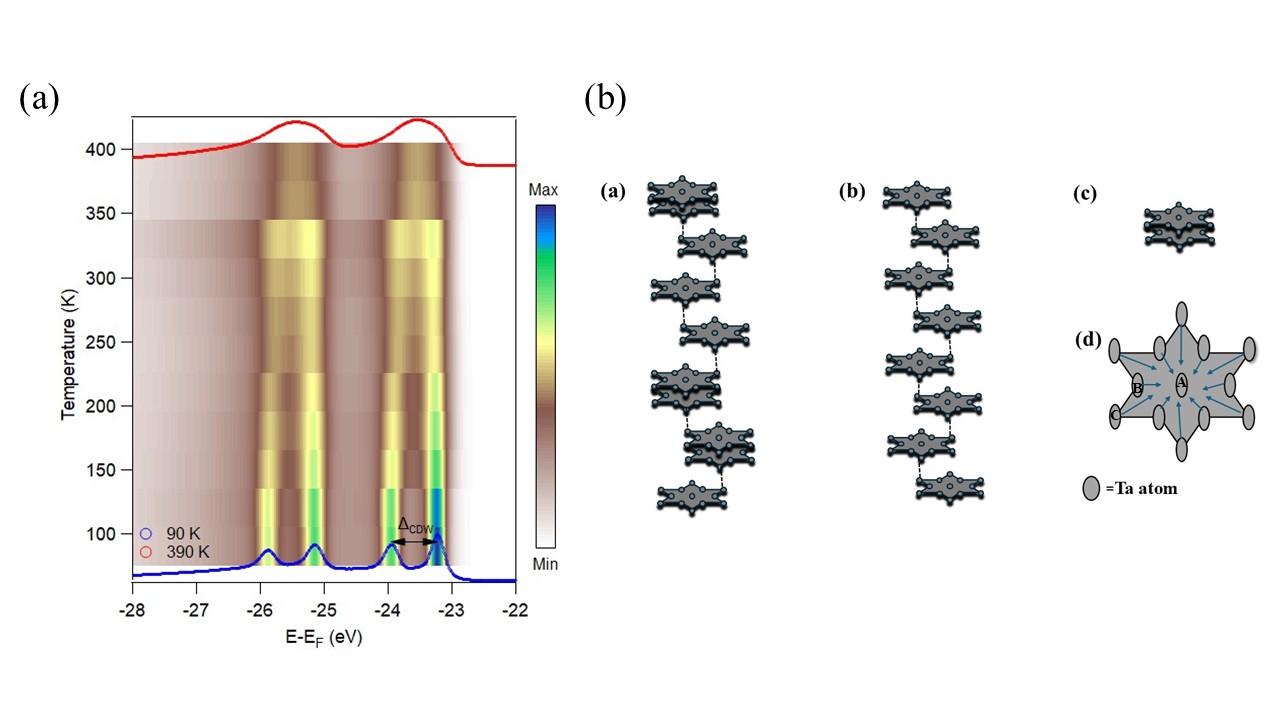}}
\caption{\textbf{Structure of 1T-TaS$_2$} \textbf{(a)} Ta 4 \emph{f} core level spectra of 1T-TaS$_{2}$ using 260 eV photon while warming up the sample from 90 K to 390 K in steps of 30 K. 
\textbf{(b)} Different out of plane stacking of Star of David structure at low temperature.  Where (i) presents the 'AL' stacking responsible for Mott insulating behaviour at low temperature. Where in the bilayers Star of David are exactly on top of each other.(ii) Represents the 'L' stacking responsible of metallic nature of the system at low temperature, where SOD's are not exactly on top of each other hence not correlated. (iii) Represents the t$_0$ bilayer formed in the 'AL' stacking where 2 layer are dimerized. (iv) Depicts the Star of David structure (SOD) where 12 Ta atoms comes inward to the 13$^{th}$ Ta atom. Due to electron density difference SOD contains 3 types of Ta atoms, central A type Ta has maximum electron density follwed by B and C type.}
\end{figure}

\begin{figure}[h!]
\centerline{\includegraphics[scale=0.65, clip]{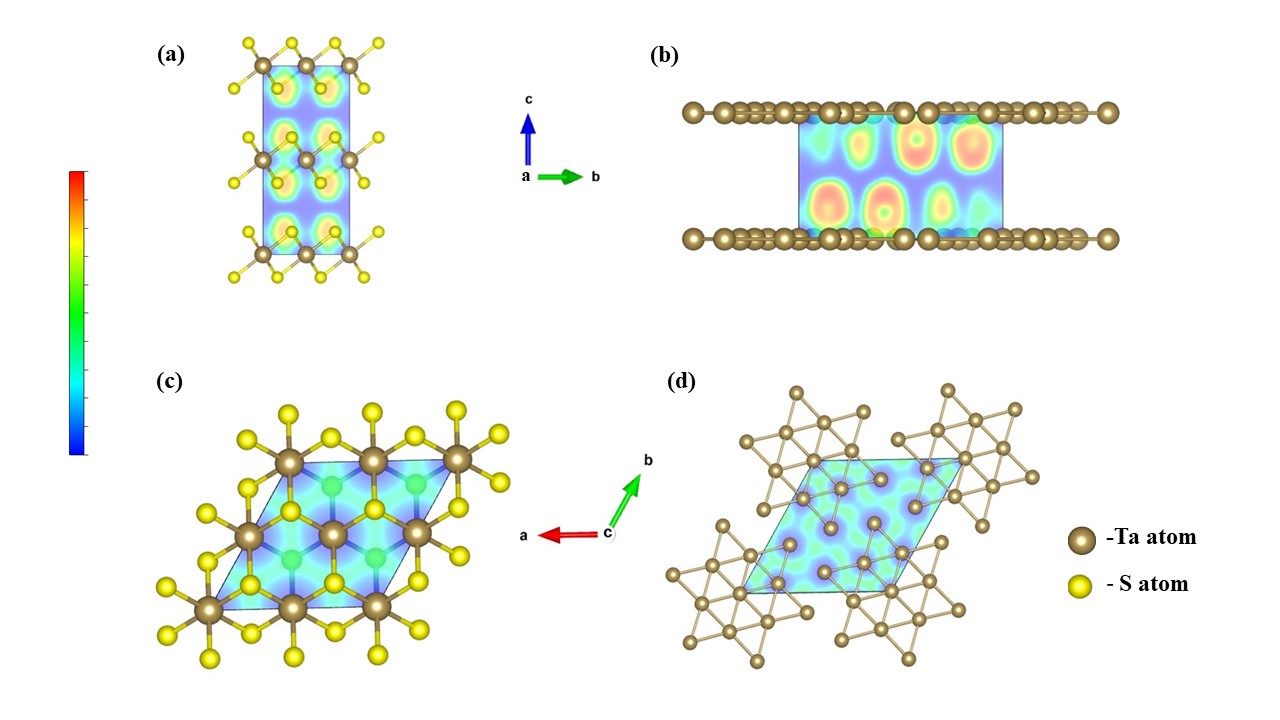}}
\caption{\textbf{Electron localization function} \textbf{a} Room temperature electron localization function calculation (ELFCAR) of the NC-CDW state along a axis. In the out of plane direction no electronic overlap exists between the Ta atoms over the Van der Waals gap. \textbf{b} Low temperature electron ELFCAR of the C-CDW (+Mott) superstructure. Substantial electronic overlap of the Ta atoms across the Van der Waals gap was found. \textbf{c} Room temperature ELFCAR of NC-CDW state observed along c axis, no electronic overlap between the Ta atoms found in the in-plane direction. Low temperature metallic state is actually the supercooled NC-CDW state, having close simillarity with the NC-CDW structure. \textbf{d} ELFCAR of low temperature C-CDW supercell structure, substantial electronic overlap between the B and C type of Ta atoms of the Star of David (SOD) structure, justifying the Mott insulating nature at low temperature.}
\end{figure}

\begin{figure}[h!]
\centerline{\includegraphics[scale=0.65, clip]{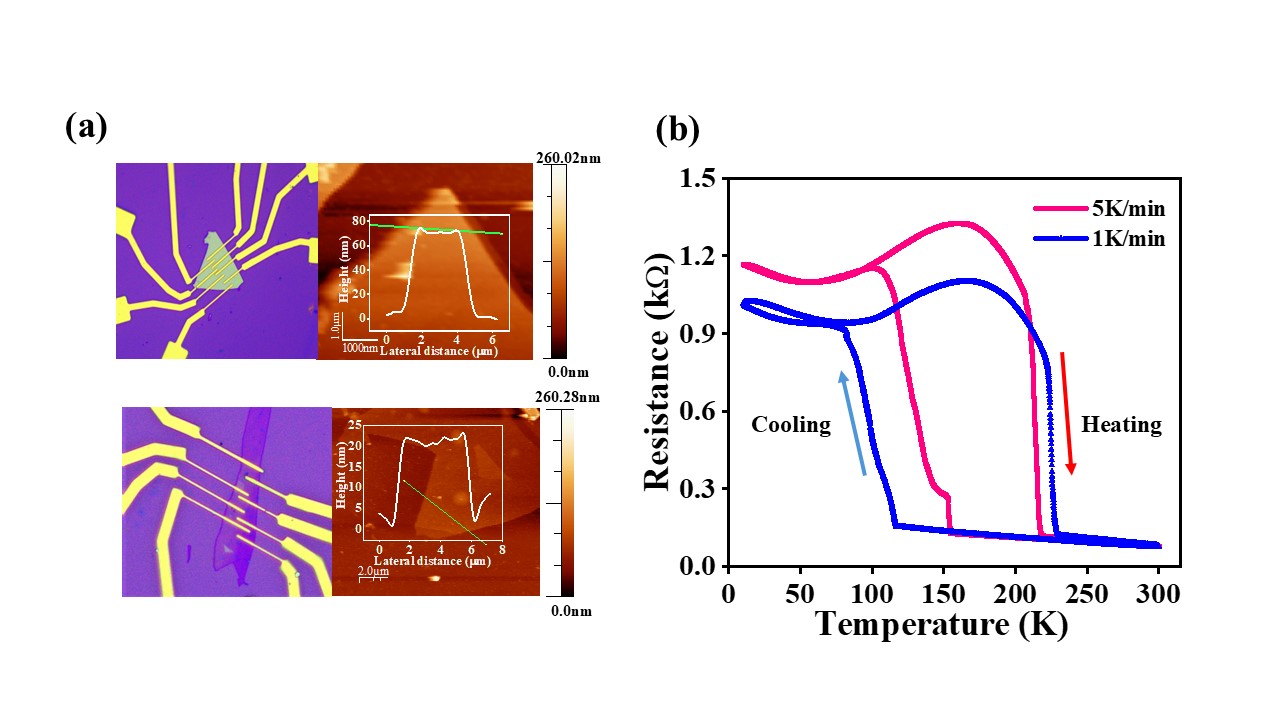}}
\caption 
{\textbf{(a)} Atomic force microscopic images, thickness of two different flakes, both exhibited ML state at low temperature doing away with the critical thickness argument for emergence of metallicity. \textbf{(b)}  Resistance (\( R \)) vs temperature (\( T \)) characteristics of the Mott Insulator (MI) state measured across the temperature range of 10 K to 300 K for two different temperature sweeping rates (1 K/min and 5 K/min). Heating-cooling cycle are indicated using arrows.}
\end{figure}
\begin{figure}[h!]
\centerline{\includegraphics[scale=0.65, clip]{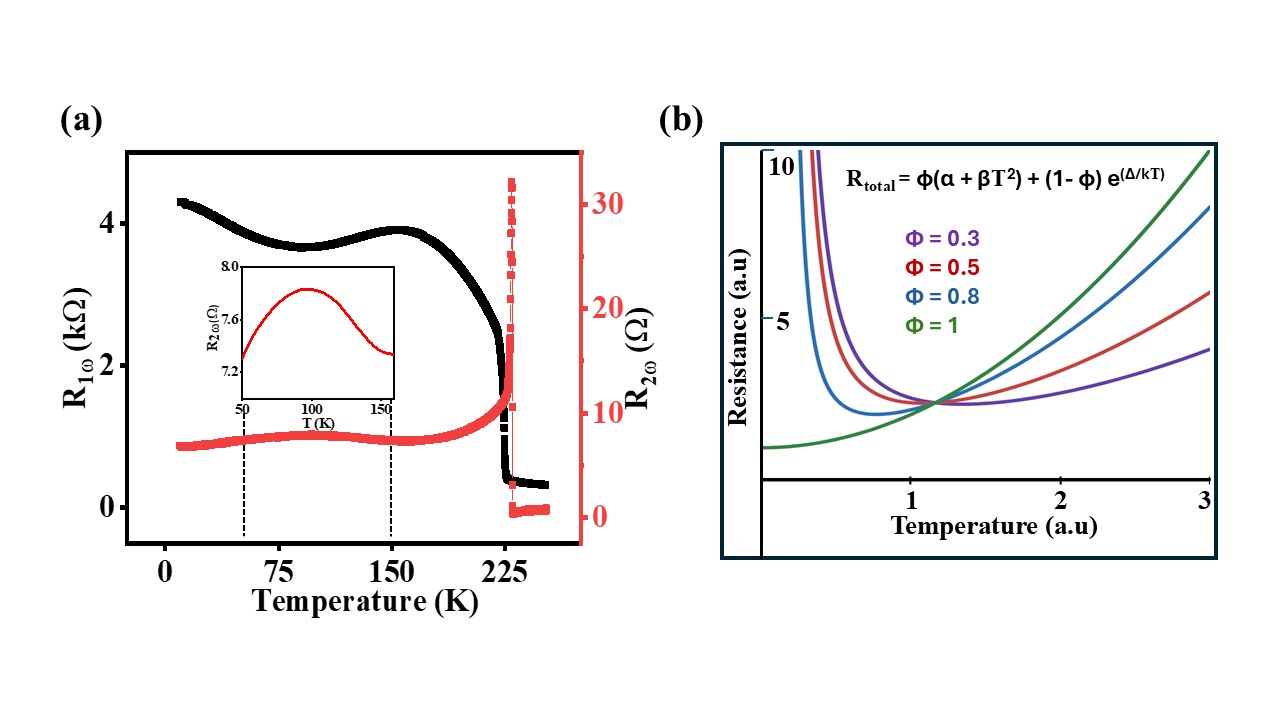}}
\caption {\textbf{(a)} \textbf{Second harmonic (2$\omega$ response of resistance with temperature of a Mott device.} Depicts the second harmonic response of resistance of a Mott device for the heating cycle, shows a sharp peak at 225 K corresponding to C-CDW to NC-CDW transition observed at 225 K. A zoomed in version of the 2$\omega$ in the temperature range 50 K to 150 K has been presented in the inset, it shows a Gaussian like peak suggesting another structural reorganisation in that zone. \textbf{(b)} Simulated resistance (R) vs temperature(T) relationship for different fractions of metallic contribution ($\phi$), where the total resistance is considered as a combination of both metallic and Mott fraction.}

\end{figure}

\begin{figure}[h!]
\centerline{\includegraphics[scale=0.65, clip]{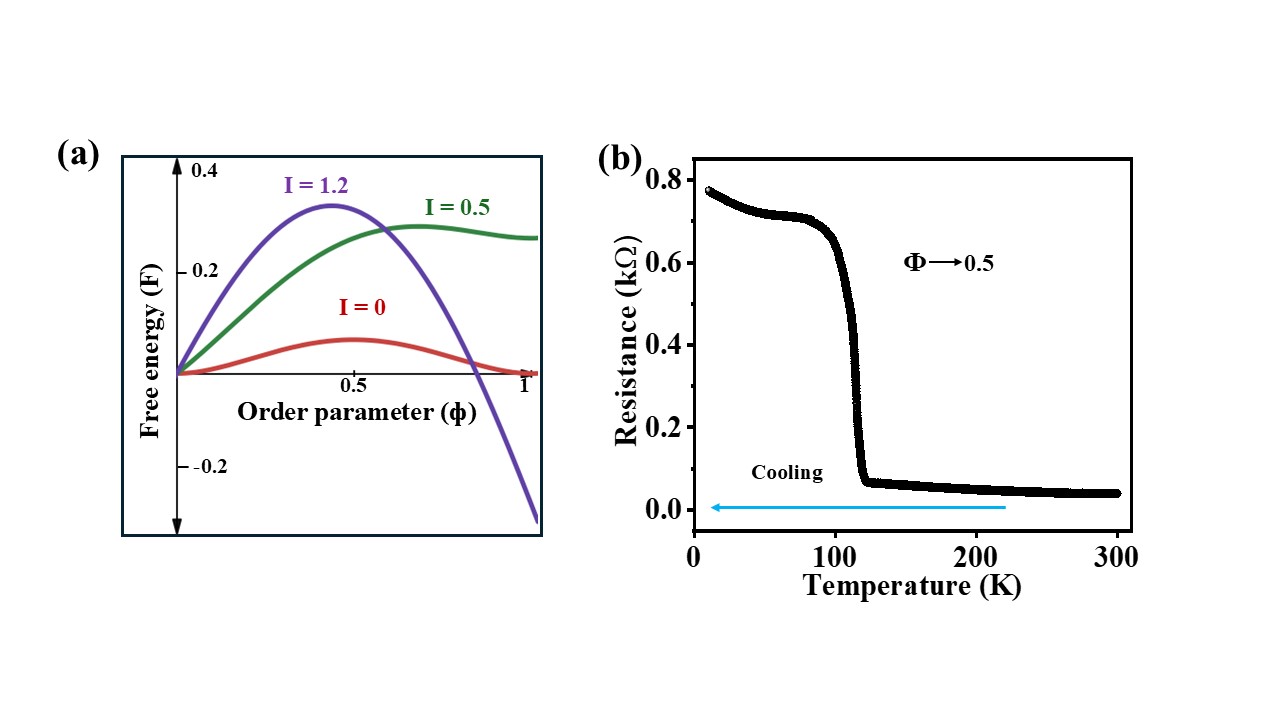}}
\caption{\textbf{(a)}{ Presents free energy (F) vs order parameter ($\phi$)for different value of current parameter, shows upto I = 0.5, metallic state becomes unstable, beyond 0.5 metallic state becomes more stable as the energy becomes negative, explaining the sudden increase in resistance at 20 $\mu$A for 10 K observed in Fig.S5(a) due to low current sweep range.} \textbf{(b)}{Presents resistance vs temperature in the cooling cycle. 800 $\Omega$ resistance and Mott-like feature at the low temperature indicates order parameter near 0.5, hence current-driven transition towards metallic state becomes feasible.}}
\end{figure}
\begin{figure}[h!]
\centerline{\includegraphics[scale=0.65, clip]{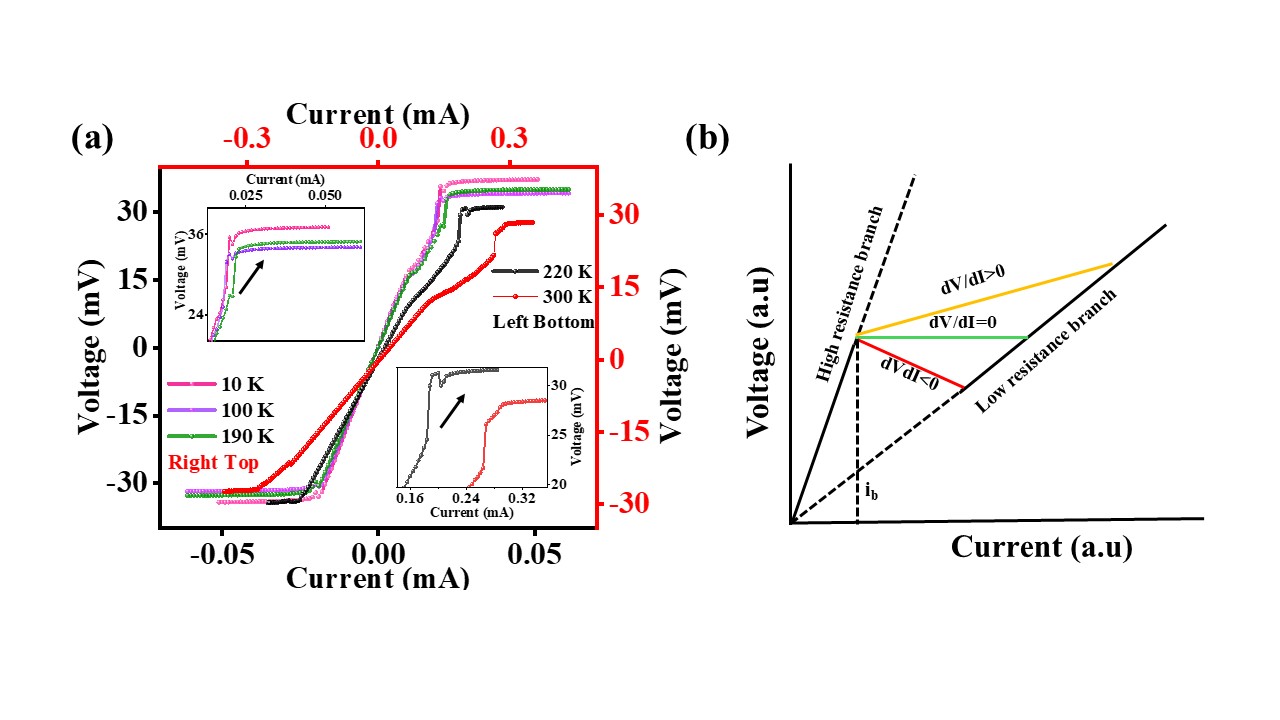}}
\caption{\textbf{(a)} \textbf{(a)} Voltage (\( V \)) vs current (\( I \)) characteristics of the Mott Insulator (MI) state measured across the temperature range of 10 K to 300 K. The data demonstrate minimal current-induced resistance changes with a sudden increase in resistance just before the change, with asymmetric behaviour in positive and negative current sweep direction.  \textbf{(b)} Represents a schematic of the evolution of \( V \)-\( I \) curves across varying current ranges, illustrating the transitions and scaling effects associated with NDR observed in the Metal like state and low resistance Mott insulating state.}
\end{figure}

\begin{figure}[h!]
\centerline{\includegraphics[scale=0.65, clip]{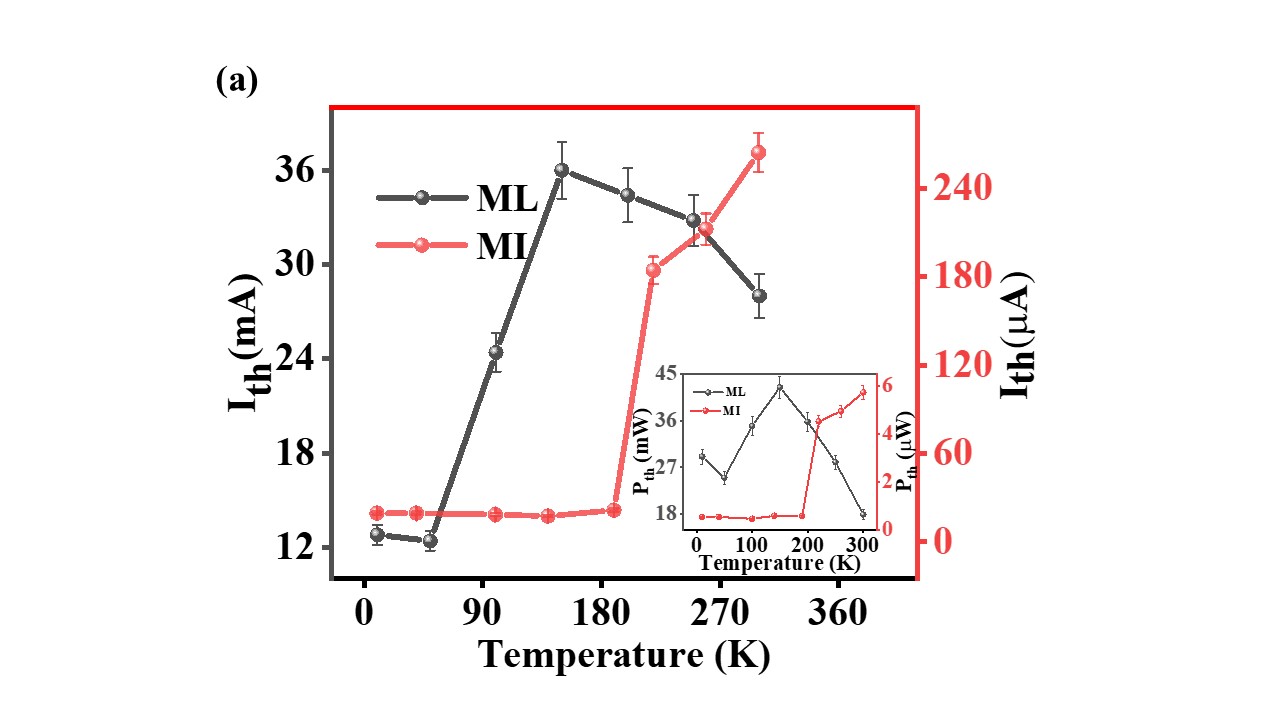}}
\caption{\textbf{Temperature Dependence of Threshold Current and Power for MI and ML States.}  \textbf{(a)} Threshold current (\( I_{\text{th}} \)) at the transition point as a function of temperature for the Metal-like (ML) and Mott Insulator (MI) states, represented by black and red curves, respectively. The ML state exhibits a steady increase in \( I_{\text{th}} \) with temperature up to 150 K, indicating the enhanced stability of conducting domains. In contrast, the MI state shows nearly constant \( I_{\text{th}} \) below 190 K, followed by a gradual decrease as the system transitions into a channel-dominated phase. The inset presents the threshold power (\( P_{\text{th}} \)) as a function of temperature for both states. In the ML state, \( P_{\text{th}} \) initially decreases, reflecting efficient current-induced domain fragmentation at low temperatures, and then increases linearly with temperature. For the MI state, \( P_{\text{th}} \) remains nearly constant up to 190 K but shows a pronounced rise at higher temperatures. These trends underscore the temperature-dependent percolation dynamics and domain reorganization that govern the resistive switching and negative differential resistance (NDR) in both phases.}
\end{figure}
\begin{figure}[h!]
\centerline{\includegraphics[scale=0.65, clip]{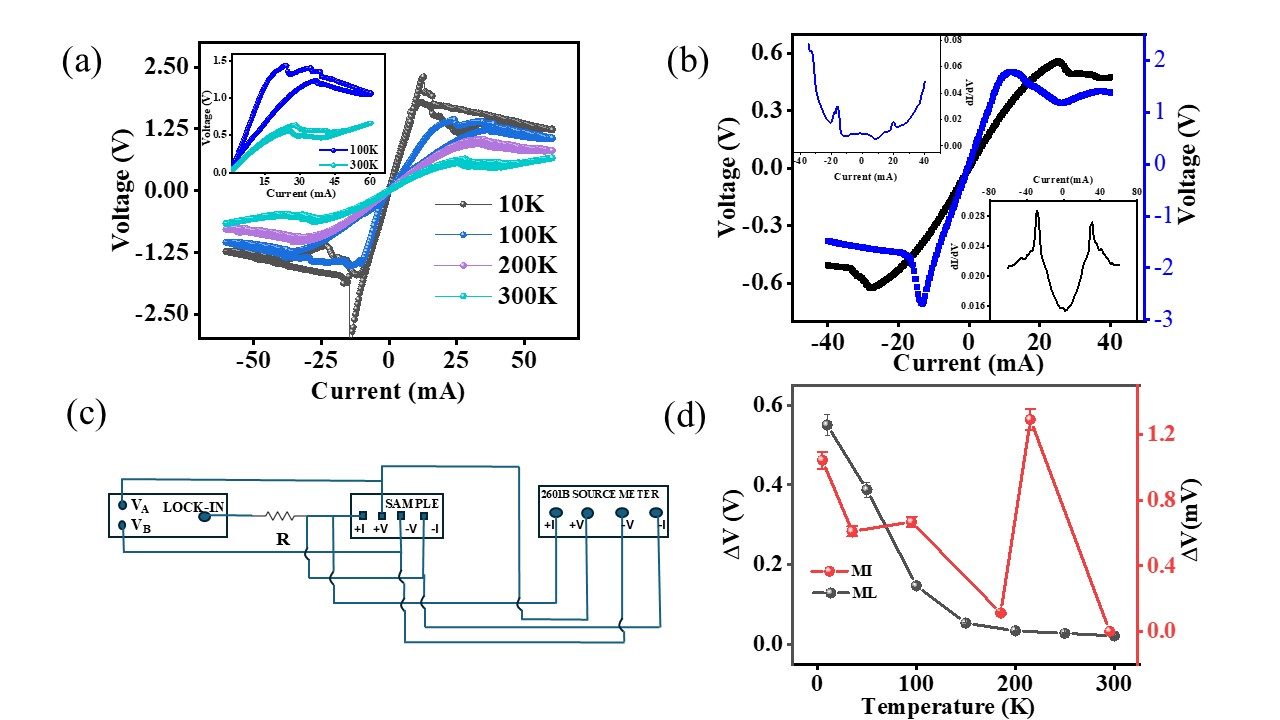}}
\caption{\textbf{(a)} Hysteresis in the voltage-current characteristics of the ML state across 10 K to 300 K. Notably hysteresis area across the NDR monotonically decreased with temperature, reflecting the sparse limit fractality behavior of the system, where fractal dimension increased from 0.37 to $\approx$ 0.9 as temperature increased from 10 K to 300 K. \textbf{(b)} Differential conductance ($dI/dV$) measurements of the ML state at 10 K and 300 K, well agreed with the V-I characteristics of the system. \textbf{(c)} Circuit configuration of the $dI/dV$ measurements. \textbf{(d)} Change of voltage across the NDR regions of ML and MI states, clearly depicts almost negligible change of voltage of the MI state, almost 1000 times more robust change in voltage for ML state depicting initial phase dependence of the NDR.}

\end{figure}
\begin{figure}[h!]
\centerline{\includegraphics[scale=0.65, clip]{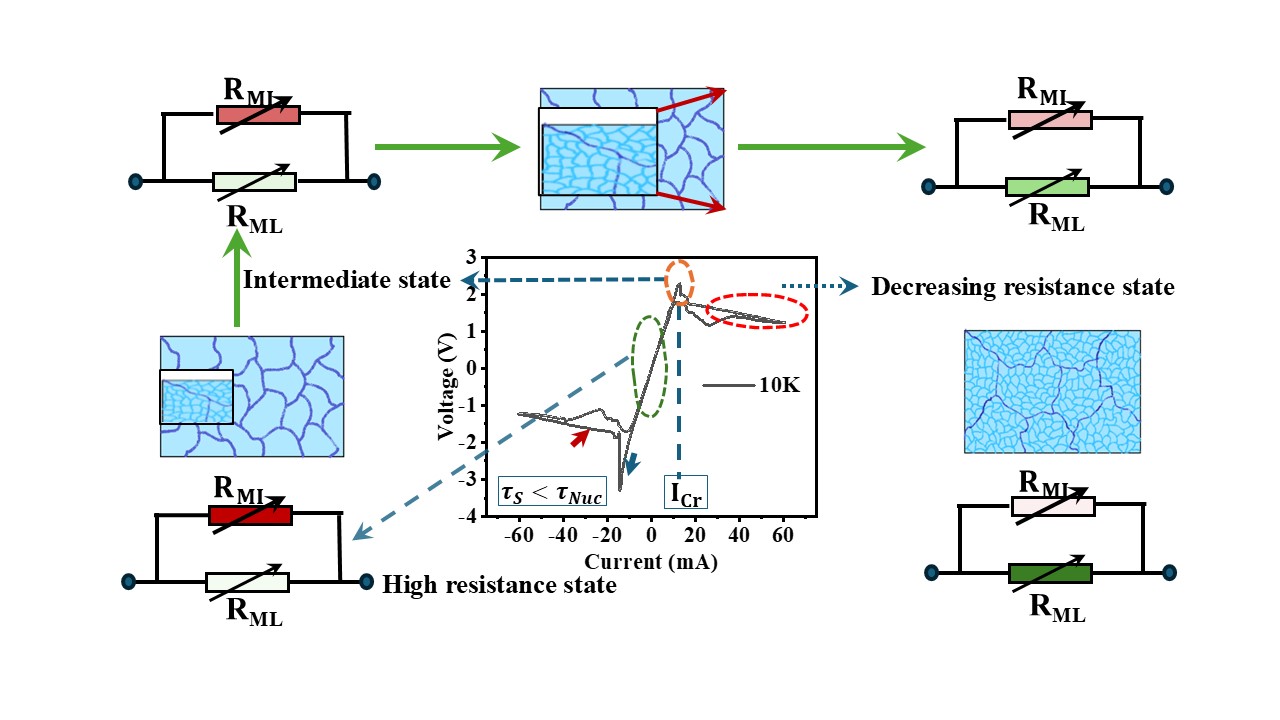}}
\caption{ Schematic of the resistor network. Interpretation of NDR and hysteresis observed in ML state V-I considering a parallel combination of resistors based on the metal like state (ML) and  Mott insulating (MI) background. The resistances (color coded with brighter  (dimmer) green (red) representing smaller (larger) resistance for the ML (MI) regions. }
\end{figure}

\begin{figure}[h!]
\centerline{\includegraphics[scale=0.65, clip]{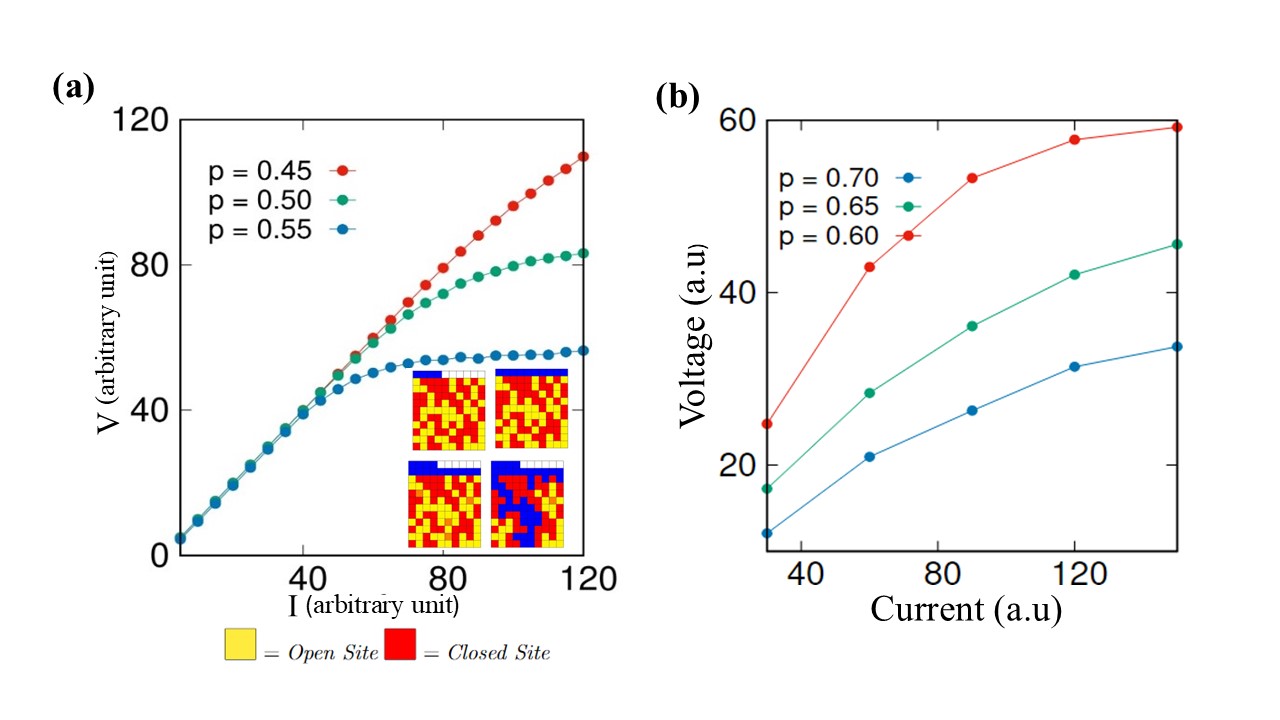}}
\caption{\textbf{(a)} Calculated \( V-I \) characteristics of the Mott state for different values of the percolation limit p. For p =0.45, much lower than the critical value 0.59 hence shows linear behaviour to some extent,as p increases towards 0.59 non-linearity appears in Voltage-Current characteristics.Inset represents checker boards, depicting site opening beyond a threshold probability due to application of current.    \textbf{(b)} Calculated \( V-I \) characteristics of the Mott state for different values of the percolation limit $>$ 0.59. For p = 0.6 non linearity is maximum beyond 0.6 V-I becomes almost linear again  representing system entered more channel (open site) dominated region hence have lower resistance.}
\end{figure}

\begin{figure}[h!]
\centerline{\includegraphics[scale=0.65, clip]{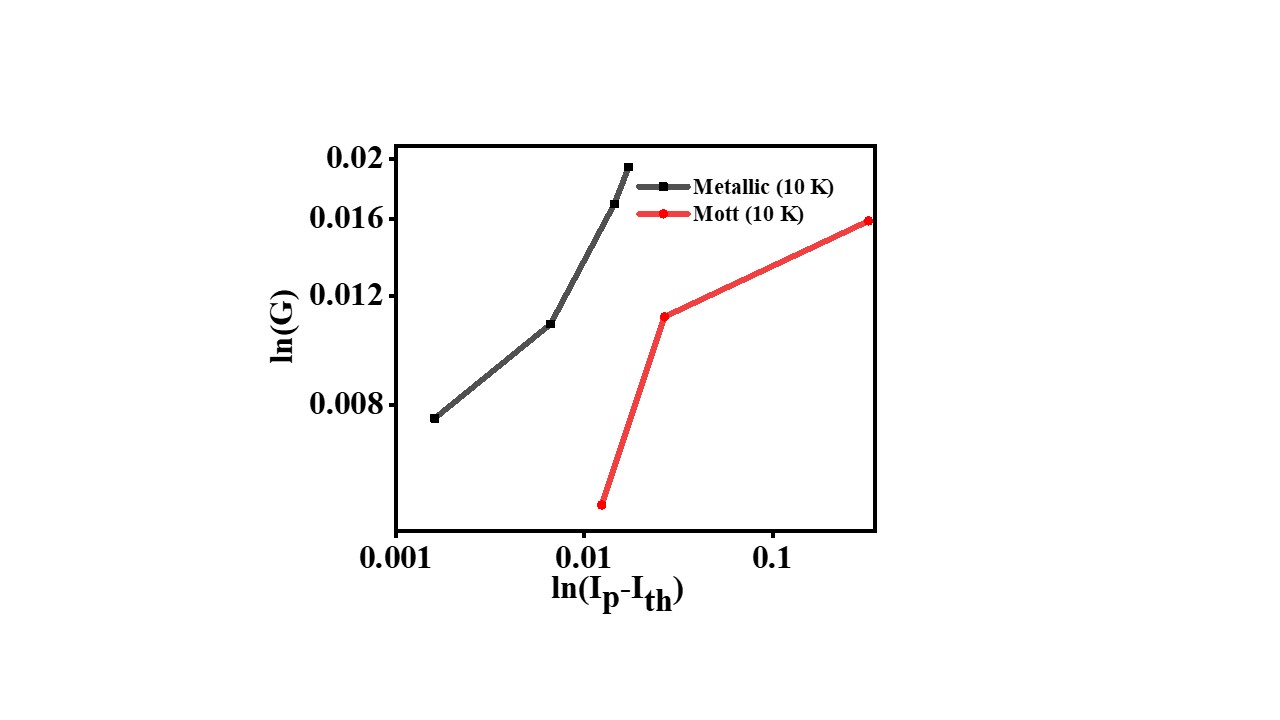}}
\caption{Depicts the logarithmic plot of conductivity (G) vs (I$_p$ - I$_th$) of both Mott (Fig. 2(a)) and metallic state (Fig. 2(b)) at 10 K, where I$_p$ is the peak current corresponding to each oscillation and I$_th$ is the critical current where NDR appeared, different value of exponent suggest multi-scale percolation dynamics in presence of fractal geometry}
\end{figure}